\documentclass[journal]{vgtc}                %

\usepackage{hyperref}

\ifpdf%
\pdfoutput=1\relax                   %
\usepackage{graphicx}                %
\DeclareGraphicsExtensions{.pdf,.png,.jpg,.jpeg} %
\else%
\usepackage{graphicx}                %
\DeclareGraphicsExtensions{.eps}     %
\fi%

\graphicspath{{figures/}{pictures/}{images/}{./}} %

\usepackage{microtype}                 %
\PassOptionsToPackage{warn}{textcomp}  %
\usepackage{textcomp}                  %
\usepackage{mathptmx}                  %
\usepackage{times}                     %
\usepackage{cite}                      %
\usepackage{booktabs}                  %

\onlineid{0}

\vgtccategory{Research}
\vgtcpapertype{theory/model}

\vgtcinsertpkg

\usepackage{array, booktabs}           
\usepackage{amsmath}
\usepackage{mathtools}
\usepackage{euscript}
\usepackage{tikz}
\usetikzlibrary{topaths,calc}
\usepackage{subcaption}
\DeclareCaptionFont{8pt}{\fontsize{8pt}{9.52pt}\selectfont}
\usepackage{todonotes}
\usepackage{mathrsfs}
\usepackage{amsmath}
\usepackage{amssymb}
\usepackage[babel]{csquotes}
\usepackage{wasysym}
\usepackage{multirow}
\usepackage{xstring}
\usepackage{colortbl}
\usepackage{framed}
\usepackage{bm}
\usepackage{multicol}
\usepackage{enumitem}

\DeclareRobustCommand{\inlinenlpsymbol}[1]{%
	\begingroup\normalfont
	\raisebox{-.2\height}{\includegraphics[height=1.5\fontcharht\font`\D]{figures/symbols/#1}}
	\endgroup
}

\newcommand\symLetter[1]{%
	\begin{tikz}[baseline=(X.base)] 
		\node (X) [color=white, fill=black, inner sep=1pt] {#1};
	\end{tikz}%
}

\newcommand{\symYes}{\CIRCLE}
\newcommand{\symNo}{\Circle}
\newcommand{\symPartial}{\LEFTcircle}
\newcommand{\symExample}{\inlinenlpsymbol{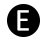}}
\newcommand{\symComparison}{\inlinenlpsymbol{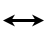}}
\newcommand{\symInterview}{\inlinenlpsymbol{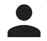}}

\definecolor{ColorFrameworkData}{HTML}{4E91D8}
\definecolor{ColorFrameworkModel}{HTML}{76AF1B}
\definecolor{ColorFrameworkVisualization}{HTML}{E40001}
\definecolor{ColorFrameworkKnowledge}{HTML}{ED8901}

\definecolor{ColorBGFrameworkData}{HTML}{EDF4FB}
\definecolor{ColorBGFrameworkModel}{HTML}{F1F7E8}
\definecolor{ColorBGFrameworkVisualization}{HTML}{FCE5E5}
\definecolor{ColorBGFrameworkKnowledge}{HTML}{FDF3E5}

\definecolor{ColorInfoBoxTitle}{HTML}{DF7575}
\definecolor{ColorInfoBoxBody}{HTML}{F8E0E0}
\definecolor{ColorInfoBoxNote}{HTML}{FAF0F0}

\definecolor{ColorInfoBoxGrayTitle}{HTML}{606060}
\definecolor{ColorInfoBoxGrayBody}{HTML}{F0F0F0}
\definecolor{ColorInfoBoxGrayNote}{HTML}{FAF0F0}

\newcolumntype{A}{>{\columncolor{ColorBGFrameworkData!50}}c}
\newcolumntype{B}{>{\columncolor{ColorBGFrameworkModel!50}}c}
\newcolumntype{C}{>{\columncolor{ColorBGFrameworkVisualization!50}}c}
\newcolumntype{D}{>{\columncolor{ColorBGFrameworkKnowledge!50}}c}

\DeclareRobustCommand{\matrixDT}[6]{
	\begin{tikzpicture}[scale=0.075, baseline]
		\fill[#1] (0,1) rectangle ++ (1,1);
		\fill[#2] (1,1) rectangle ++ (1,1);
		\fill[#3] (2,1) rectangle ++ (1,1);
		\fill[#4] (0,0) rectangle ++ (1,1);
		\fill[#5] (1,0) rectangle ++ (1,1);
		\fill[#6] (2,0) rectangle ++ (1,1);
		\draw[darkgray] (0,0) grid (3,2); 
	\end{tikzpicture} 
}

\DeclareRobustCommand{\matrixTT}[9]{
	\begin{tikzpicture}[scale=0.05, baseline]
		\fill[#1] (0,2) rectangle ++ (1,1);
		\fill[#2] (1,2) rectangle ++ (1,1);
		\fill[#3] (2,2) rectangle ++ (1,1);
		\fill[#4] (0,1) rectangle ++ (1,1);
		\fill[#5] (1,1) rectangle ++ (1,1);
		\fill[#6] (2,1) rectangle ++ (1,1);
		\fill[#7] (0,0) rectangle ++ (1,1);
		\fill[#8] (1,0) rectangle ++ (1,1);
		\fill[#9] (2,0) rectangle ++ (1,1);
		\draw[darkgray] (0,0) grid (3,3); 
	\end{tikzpicture} 
}

\DeclareRobustCommand{\inlinesymbol}[1]{%
	\begingroup\normalfont
	\raisebox{-.2\height}{\includegraphics[height=1.5\fontcharht\font`\D]{#1}}
	\endgroup
}

\DeclareRobustCommand{\inlinesymbolup}[1]{%
	\begingroup\normalfont
	\raisebox{-.05\height}{\includegraphics[height=1.3\fontcharht\font`\D]{#1}}
	\endgroup
}

\renewenvironment{leftbar}[2]
{
	\def\FrameCommand
	{%
		{\color{#1}\vrule width 3pt}%
		\hspace{0pt}%
		\fboxsep=\FrameSep\colorbox{#2}%
	}%
	\MakeFramed{\hsize\hsize\advance\hsize-\width\FrameRestore}
	\vspace*{-2mm}%
	
}
{\vspace*{-2mm}\par\unskip\endMakeFramed\vspace*{-5mm}}

\title{Communication Analysis through Visual Analytics:\\ Current Practices, Challenges, and New Frontiers}

\author{Maximilian~T.~Fischer, Frederik~L.~Dennig, Daniel~Seebacher, Daniel~A.~Keim, and~Mennatallah~El-Assady}
\authorfooter{
	\item
	M. T.\ Fischer, F. L.\ Dennig, D. Seebacher, and D. A.\ Keim are with the University of Konstanz, M. El-Assady is with the AI Center, ETH Zürich. E-mail: \{max.fischer, frederik.dennig, daniel.seebacher, keim\}@uni-konstanz.de, melassady@ai.ethz.ch.
}

\shortauthortitle{Fischer \MakeLowercase{\textit{et al.}}: Communication Analysis through Visual Analytics}

\abstract{
	The automated analysis of digital human communication data often focuses on specific aspects such as content or network structure in isolation.
	This can provide limited perspectives while making cross-methodological analyses, occurring in domains like investigative journalism, difficult.
	Communication research in psychology and the digital humanities instead stresses the importance of a holistic  approach to overcome these limiting factors.
	In this work, we conduct an extensive survey on the properties of over forty semi-automated communication analysis systems and investigate how they cover concepts described in theoretical communication research.
	From these investigations, we derive a design space and contribute a conceptual framework based on communication research, technical considerations, and the surveyed approaches.
	The framework describes the systems' properties, capabilities, and composition through a wide range of criteria organized in the dimensions (1) Data, (2) Processing and Models, (3) Visual Interface, and (4) Knowledge Generation.
	These criteria enable a formalization of digital communication analysis through visual analytics, which, we argue, is uniquely suited for this task by tackling automation complexity while leveraging domain knowledge.
	With our framework, we identify shortcomings and research challenges, such as group communication dynamics, trust and privacy considerations, and holistic approaches.
	Simultaneously, our framework supports the evaluation of systems and promotes the mutual exchange between researchers through a structured common language, laying the foundations for future research on communication analysis.
} %


\CCScatlist{ %
	\CCScat{K.6.1}{Management of Computing and Information Systems}%
	{Project and People Management}{Life Cycle};
	\CCScat{K.7.m}{The Computing Profession}{Miscellaneous}{Ethics}
}

\begin{document}
	
\firstsection{Introduction}
\label{sec:introduction}

\maketitle
Human communication has been fundamentally transformed, especially in the last two decades, becoming increasingly digital, with cost-effective, location-independent, and instant access changing communication behavior.
With this transformation to digital communication~\cite{Scolari.DigitalCommunication.2009}, new research opportunities have emerged in a wide variety of different domains, ranging from engineering to social sciences to business:
For example, it has been studied how to visualize the evolution of dynamic communication networks~\cite{Trier.DynVisCommNetworks.2008}, how discourse analysis for digital communication can be enhanced~\cite{Herring.ComputerMediatedDiscourse.2019}, or how team communication performance in business settings can be evaluated~\cite{Foltz.CommAnaTeams.2008}.
For such analyses, digital analysis methods are often used to aid and support the (semi-)manual, domain-specific research methodologies.

In this paper, we focus on the field of interactive human communication analysis and specifically on \textbf{automated and interactive communication analysis \emph{systems}} targeting written human communication (in the following: communication analysis systems), most commonly e-mails, chats, or documents.
For this paper, we define these systems as semi-automated applications that employ visual components for an interactive analysis. %
We do not consider approaches focusing primarily on a single methodology like sentiment analysis, but those that aim at a cross-methodological analysis among multiple parties.
This analysis becomes increasingly relevant in many investigative domains~\cite{Brehmer.Overview.2014, Fischer.EthicalAwarenessCommAna.2022}.

As we highlighted in previous work~\cite{Fischer.CommAID.2021}, the research into communication analysis \emph{systems} often lacks~\cite{Elzen.MultivarNetwExpl.2014, Brehmer.Overview.2014} \textbf{cross-methodological} aspects:
the majority of systems focus on either the content of communication \emph{or} on the network aspect \emph{in isolation} instead of considering the fundamental dynamics holistically.
This is in contrast to seminal works on human communication research~\cite{McLuhan.UnderstandingMedia.1964, WatzlawickBeavinJackson.Communication.1974}, recent textbooks~\cite{Pearson.HumanCommunication.2011, McLuhan.MediumIsMessage.2017}, or current communication research in psychology or digital humanities~\cite{Foltz.CommAnaTeams.2008, Mesch.SocialContextCommunicationChannels.2009}, where often -- even when digitally supported~\cite{Fan.SocialMediaAnalytics.2014, Wu.SurveyVASocialMediaData.2016} -- a holistic view is taken to consider ex- and implicit connotations in context.
In contrast, the individual analysis of content, network, and metadata can -- for interrelated tasks -- lead to an incomplete or biased view, while isolated approaches often introduce discontinuities, increase manual work and hamper cross-methodological detection.

\textbf{Existing frameworks} on digital communication analysis \emph{systems} do not adequately cover these issue due to four reasons:
First, the need for such a revised formalization has been recognized~\cite{vanAtteveldt.CommunicationComputation.2018} in communication sciences.
So far the opportunities, challenges, and pitfalls have primarily been described from an application domain-oriented perspective~\cite{Fan.SocialMediaAnalytics.2014, Wu.SurveyVASocialMediaData.2016} in the social sciences, while a systematic description is missing, only available for social-media-based approaches~\cite{Fan.SocialMediaAnalytics.2014, Wu.SurveyVASocialMediaData.2016}.
Ethical considerations~\cite{Fischer.EthicalAwarenessCommAna.2022} so far play only a small role in the system design.
Second, recent efforts have begun to map digital communication systems as a whole~\cite{Flensburg.DigitalCommunicationSystems.2020}, with a focus on content, infrastructure, and policy aspects, but leaving out the technical considerations, like methods, interfaces, and interaction concepts.
Third, the same is true for the classical communication analysis research~\cite{McLuhan.UnderstandingMedia.1964,WatzlawickBeavinJackson.Communication.1974, Pearson.HumanCommunication.2011}%
, which lacks technical considerations and is primed for analog but not digital communication.
Fourth, digital communication has also transformed the way we communicate and the modalities we use~\cite{Onyeator.HumanCommDigital.2019}, like shorter messages or emoji reactions, requiring an updated framework.

In this work, we want to bridge the \textbf{gap} between communication research and modern communication analysis system development. 
As evident in a few academic works~\cite{Hadjidj.ForensicEMailFramework.2009, Wu.OpinionFlow.2014, Koven.Beagle.2019, Fischer.CommAID.2021} and recent commercial systems~\cite{Nuix.DiscoverInvestigate.2020, Palantir.Gotham.2020, DataWalk.2020}, visualization and interactive user steering is a promising way~\cite{Keim.VisualAnalytics.2008, Yi.InteractionInfoVis.2007, Ghani.VAMultimodalSNA.2013, Sacha.KGM.2014, Herring.InteractiveMultimodalComm.2015, Wu.SurveyVASocialMediaData.2016, Conlen.DesignPrinciplesVA.2018} to begin to tackle the gap between different analysis modalities.
Lack of a common description from both a technical perspective and psychological communication research has made the systematic exploration of the field difficult.
This also prevented a broader review of how visual analytics principles are -- and could be -- employed in communication analysis, how such systems can be categorized, and what a relevant taxonomy would look like. %
The main objective of this work is to explore these systems from a primarily capability-oriented perspective, in terms of communication research, technical state of the art, and human factors.
While we consider and point out these human factors and ethical considerations as much as possible within this framing, we also refer to our accompanying paper~\cite{Fischer.EthicalAwarenessCommAna.2022} for a more detailed background and in-depth discussion on ethical awareness and human factors in communication analysis.

As part of this work, we survey state-of-the-art approaches and investigate concepts in communication research to derive a \textbf{design space} on communication research, making the following contributions:

\begin{itemize}[topsep=0pt]
	\setlength{\itemsep}{0pt}
	\setlength{\parskip}{0.06em}
	\setlength{\parsep}{0pt}
	\item The creation of a \textbf{conceptual framework} (see Figure~\ref{fig:design_space}) of communication analysis systems, based on communication research, technical considerations, and a systematic review.
	\item A state-of-the-art \textbf{survey} and comparison of existing approaches, assessing their maturity and coverage (see Table~\ref{tab:survey} and the \textbf{interactive browser} at \href{https://communication-analysis.dbvis.de}{https://communication-analysis.dbvis.de})
	\item A discussion on the open challenges and implications for future research \textbf{opportunities} on communication analysis systems.
\end{itemize}

With this contribution, we identify research challenges and  aid the comparison of approaches while creating a taxonomy for future research on communication analysis through visual analytics.

\section{Background}
\label{sec:background}
Communication analysis can use a variety of different techniques to analyze communication behavior in its entirety.
The origins of \textbf{communication research} can be traced backed to rhetoric and oratory in ancient times. %
However, as we described in previous work~\cite{Fischer.CommAID.2021}, the formal study of communication as a process started in the early 20th-century on group formation (Simmel), human nature of communication (Cooley), disparity of expression (Lippmann), and interpersonal connections (Moreno).
The works were extended to communication patterns inside groups (Bevelas~\cite{Bavelas.CommPatterns.1950} and Leavitt~\cite{Leavitt.CommPatternsGroup.1951}) as well as to computer-aided modeling (Shannon~\cite{Shannon.TheoryCommunication.1949} and Savage and Deutsch~\cite{SavageDeutsch.TransactionFlows.1960}).
From the 1960s onward, the seminal works on media (McLuhan~\cite{McLuhan.UnderstandingMedia.1964}), communication theory (Watzlawick et al.~\cite{WatzlawickBeavinJackson.Communication.1974}), inter-human communication (von Thun~\cite{SchulzvonThun.MiteinaderReden.1981}), and communication networks (Roger~\cite{Rogers.CommNetworks.1980})  established the field.
It now encompassing diverse techniques~\cite{Pearson.HumanCommunication.2011, McLuhan.MediumIsMessage.2017}, from natural language processing over social network analysis to metadata exploration.
All focus on the \textbf{human factors} and \textbf{communication context}, with channels (Watzlawick et al.) / medium (McLuhan) forming an essential analysis aspect:
Be it phrasing or omissions, narrow- or broadcasting to target audiences, subtle implicit messages, expectations of confidentiality (and frankness), or effort in crafting the messages.

With the \textbf{advent of digital} processing the laboursome manual analysis~\cite{Gerbner.CommunicationContent.1969} shifted first to digitally supported~\cite{Lombard.ContentAnalysisMassComm.2002} and later to highly automated analysis. 
However, it is noticeable~\cite{Fischer.CommAID.2021} that the analysis' completeness developed counter to automation level, with increasing specialization and isolation.
For example, modern systems can analyze communication behavior using centrality measures~\cite{Luo.SNACommChar.2015} or describe networks ties in social sciences~\cite{Borgatti.NetwAnalysisSocialSciences.2009}.
Specialized visual toolkits have been developed to analyze such networks, like Pajek~\cite{Batagelj.Pajek.1998} or Gephi~\cite{Bastian.Gephi.2009}.
However, all these approaches primarily focus on the network aspects, omitting most of the meta-data and especially the content.
Others focus on content instead, like keyword-based searches~\cite{Yoon.TextMiningPatentNetwork.2004} to filter communication or aiming to improve the understanding of communications meaning' through sentiment analysis~\cite{Pang.OpinionSentiment.2008} or topic modeling~\cite{Rehurek.TopicModelling.2010}.

However, \textbf{Visual Analytics} could support a more comprehensive analysis, as we outline in our previous work~\cite{Fischer.CommAID.2021}, discussing the potentials for holistic systems.
Even existing visual approaches often follow insular approaches, like using node-link-diagrams (e.g., Gephi~\cite{Bastian.Gephi.2009}, and many commercial solutions like IBM's i2 Analyst's Notebook~\cite{IBM.AnalystsNotebook}, Pajek~\cite{Batagelj.Pajek.1998}, Palantir Gotham~\cite{Palantir.Gotham.2020}, DataWalk~\cite{DataWalk.2020}, and Nuix Discover and Nuix Investigate~\cite{Nuix.DiscoverInvestigate.2020}).
Another class of approaches uses matrix-based approaches to analyze the communication (or social) relations, for example, MatrixExplorer ~\cite{Henry.MatrixExplorer.2006} or NodeTrix~\cite{Henry.NodeTrix.2007}.
Another set of approaches uses timeline designs like CloudLines~\cite{Krstajic.CloudLines.2011}, while others like Fu et al.~\cite{Fu.VisEmailNetw.2007} modify graph presentations through multiple planes.

The complexity and ambiguity of the exchanges and modalities~\cite{Jensen.CommunicationModality.2000} make complete automation difficult, simultaneously raising serious ethical and privacy considerations~\cite{Fischer.EthicalAwarenessCommAna.2022}.
As such, \textbf{visual analytics}~\cite{Keim.VisualAnalytics.2008}, which describes the concept of combining computational data analysis with interactive human sense-making for knowledge generation by leveraging multi-faceted data visualizations through rapid feedback loops, is uniquely suited~\cite{Fischer.CommAID.2021, Fischer.EthicalAwarenessCommAna.2022} for an holistic approach to communication analysis, considering the subtleties of human communication.

\section{Methodology}
\label{sec:methodology}
In the following, we aim towards tackling the central question of a \textbf{common description} of communication analysis:
\textit{How can the different approaches in communication analysis systems be described within a common, conceptual framework to allow their mutual comparison?}

\textbf{Framework Basis}  --- %
We propose to base such a framework on three areas of consideration:
(1) The \textbf{existing research landscape} of interactive communication analysis systems provides a foundation for the classification of approaches based on measures such as analytical goals, visualization and interaction methods, or the power of the knowledge generation process.
(2) \textbf{Communication research} offers decades of research on the particularities of (often non-technical) communication analysis. 
For this work, we consider concepts from seminal and more recent summary works~\cite{Shannon.TheoryCommunication.1949, McLuhan.UnderstandingMedia.1964,WatzlawickBeavinJackson.Communication.1974, SchulzvonThun.MiteinaderReden.1981, Mesch.SocialContextCommunicationChannels.2009, Pearson.HumanCommunication.2011, McLuhan.MediumIsMessage.2017, Conlen.DesignPrinciplesVA.2018}, described in Section~\ref{sec:background}.
Additionally, we study relevant theoretical (non-system) works in computer science, like a survey on text visualization~\cite{Kucher.SurveyTextVis.2015}, group discourse and role analysis~\cite{Hoque.ConVisIT.2015, Hoque.MultiConVis.2016, Fu.VisForum.2018, Liu.Biosignals.2021}, %
as well as works dealing with semi-manual approaches and user studies, including the human factors (e.g.,~\cite{Jagdish.LiveBehavior.2010, Forghani.CommPatternsGrandparents.2013, Brehmer.Overview.2014, Sacha.UncertaintyAwarenessTrust.2016, Gao.MultiLanguageTeams.2017, Mastrianni.TeamCommunication.2020, Correll.EthicalDimensionsVis.2019, Fischer.EthicalAwarenessCommAna.2022}).
However, many of these works miss the transfer from a theoretically analyzed concept to an actual system implementation.
(3) \textbf{Technical considerations} of the approaches, taking into account design properties such as analyzable data types, data representation, and flow, or limitations like scalability from a technical standpoint.
We discuss the findings from considerations (2) and (3) later in Section~\ref{sec:design_space}, while (1) requires a broad review:

\textbf{Existing Research Landscape -- Seed Papers}  --- %
To analyze the state of the art and contribute one angle of classification criteria, we start with a keyword-based seed literature survey.
As we aim to inform about the most common ideas in \emph{visual analysis} applications, we restrict the search to high-quality journals and conferences: IEEE Trans. of Vis. and Comp. Graph. (\textbf{TVCG} and \textbf{IEEE VIS}), Comp. Graph. Forum (\textbf{EuroVis}), Proc. of the \textbf{CHI} Conference on Human Factors in Comp. Sys. and their co-located events.
We focus on more recent solutions than can leverage technological advances in the last 15 year (i.e. beginning in mid 2007).
We are aware that thereby we exclude some useful techniques (e.g.,~\cite{Fu.VisEmailNetw.2007}).
Outside the VIS community, in journals like Digital Investigation, only few visualization approaches (e.g., E-Mail Forensics~\cite{Hadjidj.ForensicEMailFramework.2009}) %
have been published.
However, due to the absence of novel visualizations or integrations, they were not included.

\textbf{Selection Methodology}  --- %
For the actual paper selection methodology, we follow a four-step approach (see also Table~\ref{tab:paper_selection}).
First, we conducted a keyword-based seed search for the words \textit{communication} and \textit{analysis} on the titles, abstract, index terms, and contents of publications in each of the venues described above.
Secondly, we went through all these papers' titles and abstracts manually, discarding those which clearly are not concerned with communication analysis systems, reducing the selection significantly.
For CHI, the high number of initial approaches, and the high discard rate is due to the abundant use of the phrase \emph{communication} when referring to user actions.
In this step, we included approaches suggested by the domain experts.
Third, we manually looked at the remaining papers and decided if they indeed describe a communication analysis system (as defined above) or list it as a potential application.
In the final step, we validated the results by checking borderline cases, consequently removing seven papers.
Or final collection includes 41 approaches.

\textbf{Domain Expert Consultation}  --- %
To broaden the perspective, we consulted with eight domain experts by conducting an informal interview.
The experts belong to the field of law enforcement, working for various European law enforcement agencies, and each has extensive experience with digital investigations, including communication analysis, working in the field from ten to over 30 years.
They contributed \textbf{(1)} a collection of six \textbf{approaches used in practice}~\cite{Batagelj.Pajek.1998, Bastian.Gephi.2009, IBM.AnalystsNotebook, Palantir.Gotham.2020, DataWalk.2020, Nuix.DiscoverInvestigate.2020} (including commercial) as well as \textbf{(2)} insights on \textbf{their analytical needs} and \textbf{perceived challenges}.
The proposed approaches were included in the Selection Methodology from Step 2 onward and the analytical needs were considered for the classification criteria in Section~\ref{sec:design_space}.
Further, we recruited \textbf{additional domain experts} for an interview to \textbf{evaluate the completed conceptual framework}, which we discuss in Section~\ref{sec:expert_validation}.

In terms of challenges, the experts consider it unlikely an autonomous system can completely replace an experienced-saturated investigator with years of domain-specific knowledge~\cite{Fischer.EthicalAwarenessCommAna.2022} except in the narrowest or specialized of tasks.
As soon as incomplete information is involved and decisions under uncertainty have to be taken, the analysts often follow their hunches, exploring different options, but having difficulty in articulating their reasoning~\cite{Fischer.CommAID.2021}.
They explore related and connected information, which they consider important for contextual information~\cite{Fischer.CommAID.2021}.
As such, they are used to - and strongly prefer - visually-interactive tools for investigations, as it supports their understanding through rapid-feedback mechanisms~\cite{Fischer.CommAID.2021}, increasing their trust~\cite{Fischer.EthicalAwarenessCommAna.2022}.
Such systems have been increasingly deployed in fields such as investigative journalism or criminal investigations~\cite{Fischer.EthicalAwarenessCommAna.2022}.
Nevertheless, many experts are open to new developments and consider systems their companions, supporting them without patronizing or limiting them~\cite{Fischer.EthicalAwarenessCommAna.2022}, relieving them of labor-some manual work.
However, they have to be developed with analysts in mind~\cite{Fischer.EthicalAwarenessCommAna.2022}, otherwise potentially overwhelming them or missing key functionality~\cite{Fischer.HyperMatrix.2020}. %
Black box AI models are received critically except for hints, as the domain experts are often no AI experts, lacking opacity and making it difficult to prove provenance and a chain-of-reasoning that fulfills moral or legal obligations~\cite{Fischer.EthicalAwarenessCommAna.2022}.
Developing systems fulfilling these requirements while leveraging reliable XAI methods are a key challenges.

\begin{table}
	\centering%
	\caption{\textbf{Paper collection and coding process} steps: (1) Automated filtering, (2) manual filtering, (3) manual coding, (4) manual validation. The publications per venue and papers counts are detailed below.}
	\vspace{-0.2cm}
	\includegraphics[width=.95\linewidth]{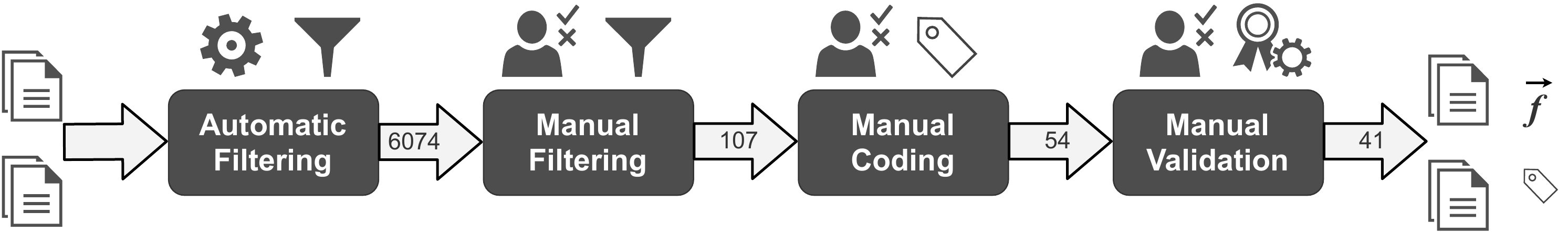} \\
	\vspace{0.1cm}
	\def\arraystretch{0.7}
	\small
	\begin{tabular}{p{3.5cm}cccc}
		
		\toprule
		Venue & \#Coll. & \#Filtered & \#Coded & \#Final \\
		\midrule[1pt]
		IEEE TVCG / IEEE VIS & 790 & 35 & 27 & 23\\ %
		Computer Graphics Forum & 495 & 17 & 11 & 8\\ %
		CHI Proceedings & 4789 & 49 & 10 & 4 \\
		Commercial Systems & - & 6 & 6 & 6 \\
		\midrule
		Total & 6074 & 107 & 54 & \textbf{41}\\
		\bottomrule%
	\end{tabular}%
	\vspace*{-0.5cm}
	\label{tab:paper_selection}
\end{table}

\section{Conceptual Framework}
\label{sec:design_space}

\begin{figure*}[!htb]
	\centering
	\includegraphics[width=\linewidth]{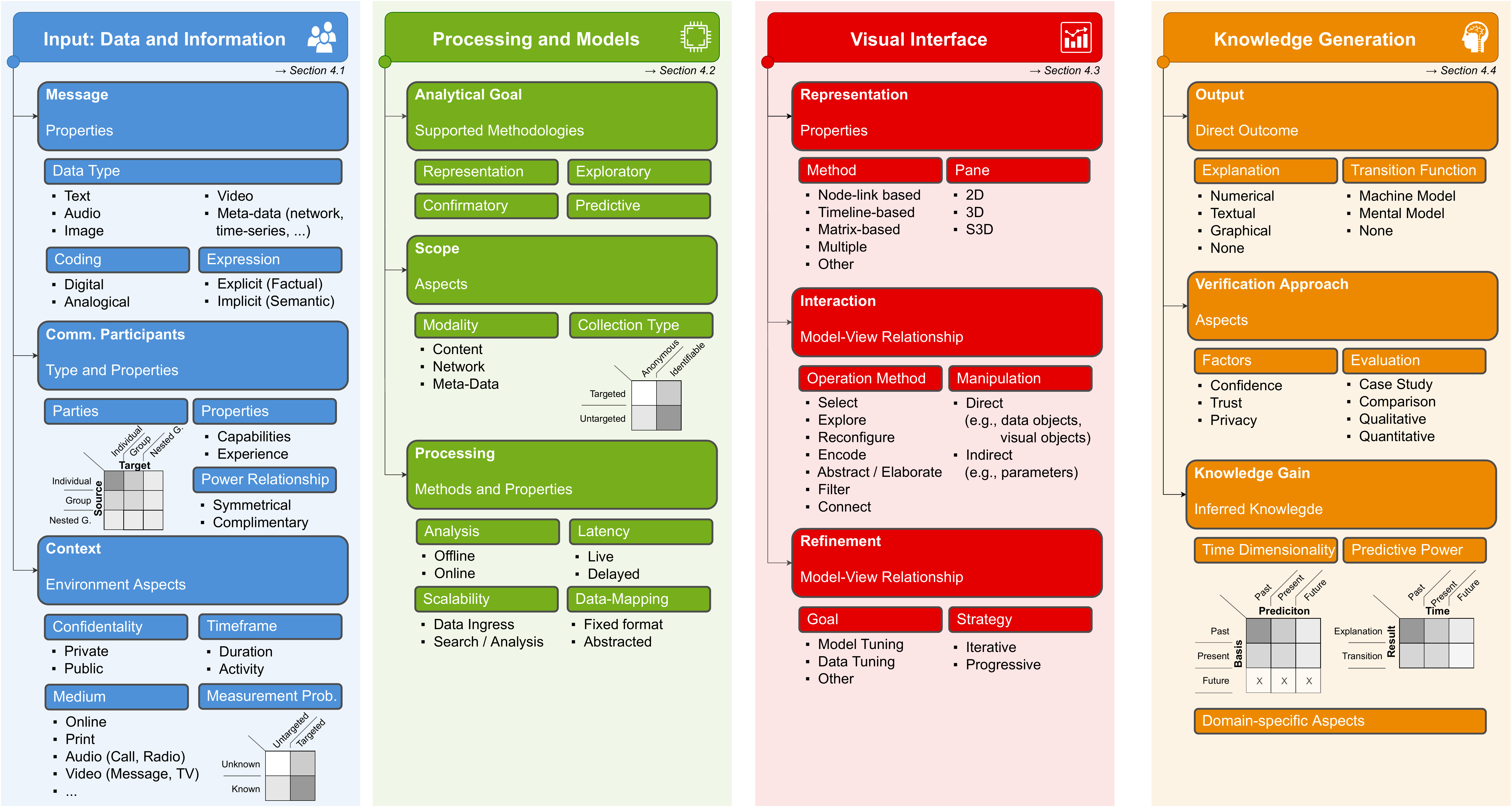}
	\caption{\textbf{Conceptual framework of communication analysis systems.}
		It consists of the four main dimensions \emph{Input: Data and Information}, \emph{Processing and Models}, \emph{Visual Interface}, and \emph{Knowledge Generation}, described in more detail in Section~\ref{sec:design_space}. %
		Each category contains several properties and sub-properties, which allow for systematical analysis of such systems.
		Note that the graphic highlights some of the most relevant aspects for individual properties, which can be used for a simplified comparison.
		However, the properties themselves are multifaceted and, for a detailed analysis, should be discussed more nuanced and in more detail than indicated by these examples.}\vspace*{-0.3cm}
	\label{fig:design_space}
	\label{fig:main_pillars} %
\end{figure*}

In the following section, we aim to construct a framework that encompasses discerning aspects of communication analysis systems.
As with any taxonomy, the framework is \emph{one} possible version of a taxonomy, developed in several iterations.
We justify our considerations overall and for each property, either based on the domain experts' requirements or referencing relevant work when applicable.
The overall framework was constructed by the authors collaboratively trough collection of aspects from the three consideration areas (see Section~\ref{sec:methodology}) and the iterative construction of a mutually exclusive and collectively exhaustive description, aiming to group related aspects and align with established nomenclatures.
We caution that many presented properties are multifaceted, and our considerations can benefit from community discussion. %
For the complete conceptual framework, see Figure~\ref{fig:design_space}, and for the full classification of the surveyed approaches, consult Table~\ref{tab:survey}.

\textbf{Main Considerations} ---
One standard \textbf{structuring methodology} is to use a task-based grouping~\cite{Conlen.DesignPrinciplesVA.2018, Fischer.HyperMatrix.2020}.
However, sometimes very different methods are employed for the same task:
for example, for discovering key persons in a communication network, SNA-based~\cite{Ghani.VAMultimodalSNA.2013} approaches using centrality measures and node-link visualizations are equally applicable as interactively visualized geometric deep learning models~\cite{Fischer.HyperMatrix.2020}. %
However, both methods have very different side effects, visualization, and interaction techniques and make very different assumptions on the data.
Instead, we follow the second large methodology and use a property, representation, and methods-based taxonomy~\cite{Kucher.SurveyTextVis.2015}.

As our \textbf{primary goal} is to design a conceptual framework of communication analysis through \emph{visual analytics}, we motivate the main areas by Keim et al.'s established process model~\cite{Keim.VisualAnalytics.2008}, but develop each of the four areas specifically for communication analysis using considerations from communication research and our survey.
We chose Keim et al.'s model because it is considered the most widely used visual analytics model~\cite{Sacha.KGM.2014}, compared to similar models like Green's Human Congnition Model~\cite{Green.VAHumanCognitionModel.2008} or van Wijk's Visualizaton Model~\cite{vanWijk.VisualizationModel.2005}.
Further, the differences between these models are slim from a compartmentalizing perspective, having similar dimensions.
We slightly modify Keim et al.'s terminology, proposing \textbf{four main dimensions}, characterized further in Figure~\ref{fig:main_pillars} and the following sections:
(I) Input: Data and Information (\ref{sec:design_space_pillar_1}) encompasses the (inferred) content and context with respect to communication research, (II) Processing and Models (\ref{sec:design_space_pillar_2}) discusses the analytical goals and scopes of the systems, (III) Visual Interface (\ref{sec:design_space_pillar_3}) presents visual and interaction techniques employed, and (IV) Knowledge Generation (\ref{sec:design_space_pillar_4}) discusses the information flow.

\subsection{Input: Data and Information~{\color{ColorFrameworkData}\inlinesymbol{symbols/icon_pillar_input}}}
\label{sec:design_space_pillar_1}
This category focuses on information, context, and environment of the communication, in particular theoretical aspects and data properties, with the structuring partly based on classical communication research~\cite{Shannon.TheoryCommunication.1949, McLuhan.UnderstandingMedia.1964,WatzlawickBeavinJackson.Communication.1974, SchulzvonThun.MiteinaderReden.1981, Mesch.SocialContextCommunicationChannels.2009, Pearson.HumanCommunication.2011, McLuhan.MediumIsMessage.2017, Conlen.DesignPrinciplesVA.2018}.
Therefore, we are discussing the content and meaning, context, and relationship aspects of communication extensively.
Building on established frameworks~\cite{Shannon.TheoryCommunication.1949, WatzlawickBeavinJackson.Communication.1974, SchulzvonThun.MiteinaderReden.1981}, we propose to focus on three interrelated areas: the information as \emph{message}, the \emph{communication participants}, and the \emph{environment} (or context).

\subsubsection{Message}
The \emph{message}~\cite{SchulzvonThun.MiteinaderReden.1981} (also central channel~\cite{Shannon.TheoryCommunication.1949} or content~\cite{WatzlawickBeavinJackson.Communication.1974}) refers to the entailed information.
From a system's perspective the distinction by \emph{data type} is obvious, while the information can be considered from its actually transported content (\emph{coding}~\cite{WatzlawickBeavinJackson.Communication.1974}) and its orthogonal interpretation (\emph{expression} levels~\cite{SchulzvonThun.MiteinaderReden.1981}):

\textbf{Data Type} ---  %
The data type refers to the content type from a technical point of view.
When looking at data classification in information visualization~\cite{Card.InfoVis.1999}, we can identify several data types which are relevant for communication analysis:
\emph{text}, \emph{audio}, \emph{image}, \emph{video}, and meta-data, related to \emph{network} as well as \emph{time-series}.
Based on the usage in current approaches (see Section~\ref{sec:methodology}), the two most relevant ones are text data (e.g., extracting topics from text~\cite{Cui.TextFlow.2011}) and relation network (structure) data (e.g., social graphs between communication participants~\cite{Bastian.Gephi.2009}).
However, communication can also happen via audio (e.g., telephone or VoIP) or via video chats, comprising audio and moving images, i.e., video data.
While our framework focuses primarily on this written (i.e., text) communication, we include these types for completeness.
Therefore, it would be possible to include the detection of facial expressions using deep learning~\cite{Ng.DLEmotionRecognition.2015} to analyze the analogical code and set it into context (see below).
Meta-data in the form of time-series data (e.g.,~\cite{Krstajic.CloudLines.2011}, extracting event order and relevance) is often relevant for the communication \emph{context}, for example regarding regularity and duration.

\textbf{Coding}  --- %
The transported meaning is a core aspect of the communication, which splits into spelled out (\emph{coding}) and inferred (\emph{expression}) meaning.
Based on Watzlawick et al.~\cite{WatzlawickBeavinJackson.Communication.1974}, the spelled out communication content~\cite{McLuhan.UnderstandingMedia.1964} can be regarded as coding, either in \emph{digital} or \emph{analogical} (sic!) form.
The digital code roughly refers to the actual meaning of the transmitted information in a symbolic system (e.g., the writing ``the sky is blue''), while the analogical code refers to how something is communicated, including cues (e.g., biosignals, like winking, or emoticons).
Analogical analysis is rare (e.g., message sentiment~\cite{Hoque.ConVis.2014, El-Assady.ConToViMultiPartyTopicSpaces.2016}), partly due to the information loss in digital transmissions.

\textbf{Expression}  --- %
Similar, but orthogonal to it is the \emph{expression}, which describes the intended or inferred information extracted from the content.
It can be \emph{explicit} factual information (the fact that the sky is blue) or information \emph{implicitly} contained and must be inferred, for example, from the semantics (or character~\cite{McLuhan.UnderstandingMedia.1964} of the message).
For example, the sky is blue - ``let's go hiking now''.
Code and expression together allow to classify systems by their capability to leverage both digital (e.g., actual content information) and also analogical codes (e.g., inferred as sentiment analysis) while judging support for explicit (e.g., keyword-based search like~\cite{IBM.AnalystsNotebook}) and also implicit (e.g., named entity recognition like~\cite{El-Assady.NEREx.2017}) content.
Most approaches consider the explicit level, and several, especially text-based ones, also the implicit level.

\subsubsection{Communication Participants}
Of central importance are the participants in a communication.
The \emph{scale} of the communication is determined by its audience. In correlation with the \emph{Context}, this determines different modes like narrow-casting (few, restricted participants), broad-casting (large audience), or targeting (specific participants), in turn influencing (or being influenced by the communication \emph{medium}).
In communication research~\cite{McLuhan.UnderstandingMedia.1964, WatzlawickBeavinJackson.Communication.1974} these aspects are usually considered as part of the context (see below).

\textbf{Parties} ---  %
According to the domain experts, the involved types of the parties can be a single other participant (with oneself as a special case) or different forms of groups (homogeneous groups or heterogeneous groups with subgroups) and differ between the sender and receiver sides.
Therefore, we propose to structure the approaches based on their support to analyze the communication between a source and a target in a 3x3 matrix (individual, group, nested groups), e.g., individual to individual is encoded as~\resizebox{!}{.75\baselineskip}{\matrixTT{black}{white}{white}{white}{white}{white}{white}{white}{white}} (e.g., no group support whatsoever~\cite{Cao.TargetVue.2016}).
Counterintuitively, the matrix might not always be symmetric.

\textbf{Properties} ---  %
The properties of these participant(s) can be manifold.
One possible classification can describe them along their \emph{capabilities} and their \emph{experience} (knowledge of context).

\textbf{Power Relationship}  --- %
The (power) relations between the parties have a strong influence, with differences in \emph{push} and \emph{pull}.
A possible classification~\cite{WatzlawickBeavinJackson.Communication.1974} distinguishes between \emph{symmetrical} (equal grounds) or \emph{complementary} (dependence) relations, going beyond mere directionality.
However, the relationship is rarely considered explicitly in existing research (e.g.,~\cite{Cao.Whisper.2012} analyses the changing relations inside a group during information diffusion).

\subsubsection{Context}
The context of communication is essential~\cite{Shannon.TheoryCommunication.1949, McLuhan.UnderstandingMedia.1964,WatzlawickBeavinJackson.Communication.1974}, because it strongly affects the implicit interpretation among participants.
We focus on the context of the external environment (\emph{confidentiality}, \emph{measurement}, and \emph{medium}), and of the message (\emph{timeframe}).

\textbf{Confidentiality}  --- %
The confidentiality of the communication channel can strongly influence the communication coding, for example through aversion or code-words (see also \emph{factors} in Section~\ref{sec:design_space_pillar_4} and our work~\cite{Fischer.EthicalAwarenessCommAna.2022} on human factors).

\textbf{Measurement Problem} ---  %
Closely related is the measurement problem, where the analysis interferes and influences the communication coding and expression simply due to its (possible) presence, as indicated by the domain experts.
The communication is affected by the participants' awareness, so they might adapt their behavior, use coded language, are less honest, implicit, or communicate not at all~\cite{Fischer.EthicalAwarenessCommAna.2022}.
This also concerns trust and reliability, both for parties and analysts~\cite{Correll.EthicalDimensionsVis.2019, Fischer.EthicalAwarenessCommAna.2022}.
We categorize this aspect into a quadrant, between (expected) knowledge about the analysis, which can be known or unknown to be true.

\textbf{Timeframe}  --- %
The timeframe when communication is occurring is highly relevant (e.g., for event correlation~\cite{Stoiber.netflower.2019}).
It can be described from the perspective of its duration and the activity during it~\cite{SeebacherFischer.ConversationalDynamics.2019}.

\textbf{Medium} ---  %
The communication medium~\cite{McLuhan.UnderstandingMedia.1964, SchulzvonThun.MiteinaderReden.1981} is partly covered (or mutually induced) by the participants and also the message type, coding, expression, and other contextual factors.
Nevertheless, it deserves its own spot, in particular, due to media-typical characteristics and its relevance in research~\cite{McLuhan.UnderstandingMedia.1964}.

\subsection{Processing and Models~{\color{ColorFrameworkModel}\inlinesymbol{symbols/icon_pillar_model}}}
\label{sec:design_space_pillar_2}
After defining the data and information available, we study the particularities of processing and model creation from this information.
We consider a technical perspective in visual data analysis, following the Golden Circle model by Simon Sinek~\cite{Sinek.WhyWhatHow.2009} to answer the why? (\emph{Analytical Goal}), the what? (\emph{Model Scope}), and the how? (\emph{Processing}).

\subsubsection{Analytical Goal} %
We start with why~\cite{Sinek.WhyWhatHow.2009}, categorizing them by the aim of the analysis, which determines the analytical tasks to achieve it.
We align our classification by the standard definition of analytical tasks in visual analytics~\cite{Sarikaya.VADesignFactors.2018}.
This includes the category \emph{representation} for a fixed analysis task (present existing data), \emph{confirmatory} analysis for a directed search (to validate a hypothesis about the data), and \emph{exploratory} analysis for an undirected search (find interesting anomalies in the data).
Another goal in communication analysis involves \emph{predictive} analysis (e.g., to analyze the future diffusion of information~\cite{Wu.OpinionFlow.2014}), to draw conclusions, which can also be regarded as overarching all methodologies and is related to the knowledge extracted (see Section~\ref{sec:design_space_pillar_4}).

\subsubsection{Scope}
The scope answers the what?, determining the generic capabilities \emph{on} the information.
Other scopes are defined by their data (see Section~\ref{sec:design_space_pillar_1}) and knowledge generation (see Section~\ref{sec:design_space_pillar_4}) support.

\textbf{Modality}  --- %
The analysis modality categorizes into the three~\cite{Fischer.CommAID.2021} core aspects: \emph{meta-data} (e.g., like time-series~\cite{Krstajic.CloudLines.2011}), \emph{network} (e.g., social graphs~\cite{Bastian.Gephi.2009}), and \emph{content} (e.g., conversation order~\cite{ElAssady.ThreadReconstructor.2018}).

\textbf{Collection Type} ---  %
The collection type is the logical composite to the \emph{Measurement Problem}, defining how the data was acquired and its corresponding analysis implications.
We propose to categorize it into a quadrant between targeting methodology and anonymity level (see the relevant part in Figure~\ref{fig:design_space}).
The former can be either targeted (specific communication from a restricted set of users) or untargeted (unfound bulk collection).
The latter can either be high (anonymized or pseudo-anonymized) or low (identifiable).
Different configurations might pose particular challenges to the analysis model regarding aspects such as scalability and inference capability through class imbalance or uncertainty~\cite{Fischer.EthicalAwarenessCommAna.2022}.
For example, the targeted analysis of identifiable communication participants can focus on the actual exchange and leverage context and relationship information.
The untargeted analysis of pseudo-anonymized communication instead often results in a search for the needle in the haystack and can rarely leverage background.

\subsubsection{Processing}
Due to existing heterogeneity, we focus on generic aspects: the \emph{analysis} approach and \emph{latency}, \emph{scalability} as KPI, and \emph{data-mapping} as power.

\textbf{Analysis}  --- %
The employed techniques and algorithms often differ significantly between \emph{offline} analysis and \emph{online} analysis.
Loading a dataset once would be considered the former type, while batch (e.g., updating data with changes~\cite{IBM.AnalystsNotebook} and, in particular, streaming approaches can be classed as the latter.
Most approaches only cover offline analysis.

\textbf{Latency}  --- %
The latency is orthogonal to the analysis.
Research~\cite{Pearson.HumanCommunication.2011} indicates that latency in the communication can significantly affect it, as well as its analysis.
The two primary options are (nearly) instantaneous communication, like in an active \emph{live}~\symLetter{L} chat (e.g., live monitoring and analysis~\cite{Palantir.Gotham.2020}) or \emph{delayed}~\symLetter{D} communication, such as e-mail or as a document
Differentiation into these two groups~\cite{Pearson.HumanCommunication.2011} is often enough for most differences in reaction and behavior, although the latency can play a role (e.g., answering under time pressure).

\textbf{Scalability} ---  %
The scalability of a KPI can be defined on two levels:
First, on the \emph{data-ingress} level, which defines the amount a system can import, analyze, and visualize initially.
The second aspect is the scalability on the \emph{search} and analysis side, for example, during exploratory analysis.
For example, how many results can be shown simultaneously?
We roughly categorize both aspects into few (less than ten, I), medium (order of hundredths to thousands, II), and huge (more than 10k, IIII).

\textbf{Data-Mapping}  --- %
Supporting data mapping increases the analytical power of the systems.
Supporting a flexible import system that allows mapping properties in contrast to a fixed data format is extremely important to the domain experts and often aligns with support for merging different data sources.
For example, many systems cannot load multiple datasets and combine fields like usernames but only consider a single dataset (e.g., e-mails) in a fixed format.

\begin{table*}[!ht]
	\caption{\textbf{Communication analysis system classification} summarizing the approaches properties. The classification criteria are formed on a subset based on the conceptual framework we developed in Section~\ref{sec:design_space} and which is shown in Figure~\ref{fig:design_space}.
	The selected approaches and most categories are also available on a dedicated website at \href{https://communication-analysis.dbvis.de}{communication-analysis.dbvis.de}, also permanently archived with OSF~\cite{Fischer.OSFCommunicationSurveyWebsite.2022}. For details, see Section~\ref{sec:discussion_limitations_future_work}.}
	\label{tab:survey}
	
	\scriptsize%
	\fontsize{5}{6}
	\selectfont
	\centering%
	\setlength\tabcolsep{1.20pt}
	\renewcommand{\arraystretch}{0.8}
	
	\begin{tabular}{r|l|c|AAAAA|AA|AA|A|A|A|BBBB|BBB|B|B|BB|B|CCC|CCCCCCC|CC|CC|CCC|DDD|DD|DDD|D|D|D|}
		\multicolumn{55}{>{\linespread{0.95}\scriptsize}p{\linewidth}}{%
			\vspace{-0.7cm}
			\begin{multicols}{2}
				\textbf{Generic Properties}: Approach supports \symYes~a property or not \symNo, or support is only very limited \symPartial~(e.g., show associated data without analysis).\newline%
				\textbf{$\bm{\dagger}$}: Indicates commercial systems or approaches widely used in industry.\newline%
				\textit{The following encodings have special symbols:}\newline%
				\textbf{All Matrix Symbols}: For an explicit labeling, see explanations in  Figure~\ref{fig:design_space}.\newline
				\textbf{Group Communication}: The approach supports different types of group communication analysis, as encoded in a 3x3 matrix.
				The rows (source) and columns (target) specify individual, group, and nested groups, respectively.
				Groups as source are not supported, and nested groups not at all.\newline%
				\textbf{Latency}: Support of live \symLetter{L} or a delayed \symLetter{D} analysis a posteriori.\newline%
				\textbf{Scalability}: Tens (I), hundredth to thousands (II), or millions (IIII) of cases.\newline%
				\textbf{Evaluation}: Case study \symExample, technique comparison \symComparison, and expert study \symInterview. \newline%
				\textbf{Time Dimensionality}: Support for different time dimensionalities encoded in a 2x3 matrix.
				Rows form the knowledge-basis (either past or present), and the columns specify if knowledge about the past, present, or future is inferred.
				\newline%
				\textbf{Predictive Power}: The approach's predictive power encodes in a 2x3 matrix the type of output (explanation or transition, as rows) in relation to the time dimensionalities past, present, and future, as columns.
			\end{multicols}
			
		} \\[-0.2cm]

\rotatebox{90}{}
& \rotatebox{90}{}
& \rotatebox{90}{}
& \multicolumn{5}{A|}{\rotatebox{90}{Data Type}} 
& \multicolumn{2}{A|}{\rotatebox{90}{Coding}} 
& \multicolumn{2}{A|}{\rotatebox{90}{Expression}} 
& \rotatebox{90}{Parties}
& \rotatebox{90}{Power Rel.}
& \rotatebox{90}{Measur. Pr.}
& \multicolumn{4}{B|}{\rotatebox{90}{Method.}} 
& \multicolumn{3}{B|}{\rotatebox{90}{Modality}} 
& \rotatebox{90}{Analysis}
& \rotatebox{90}{Latency}
& \multicolumn{2}{B|}{\rotatebox{90}{Scalability}} 
& \rotatebox{90}{Data-Map.}
& \multicolumn{3}{C|}{\rotatebox{90}{Pane}} 
& \multicolumn{7}{C|}{\rotatebox{90}{Operation M.}} 
& \multicolumn{2}{C|}{\rotatebox{90}{Manipulation}} 
& \multicolumn{2}{C|}{\rotatebox{90}{Goal}} 
& \multicolumn{3}{C|}{\rotatebox{90}{Strategy}} 
& \multicolumn{3}{D|}{\rotatebox{90}{Explanation}} 
& \multicolumn{2}{D|}{\rotatebox{90}{Trans. Func.}} 
& \multicolumn{3}{D|}{\rotatebox{90}{Factors}} 
& \rotatebox{90}{Evaluation}
& \rotatebox{90}{Time Dim.}
& \rotatebox{90}{Predictive P.}
\\
  \cmidrule{3-3}  \cmidrule{4-9} \cmidrule{9-11} \cmidrule{11-13} \cmidrule{13-13}  \cmidrule{14-14}  \cmidrule{15-15}  \cmidrule{16-20} \cmidrule{20-23} \cmidrule{23-23}  \cmidrule{24-24}  \cmidrule{25-27} \cmidrule{27-27}  \cmidrule{28-31} \cmidrule{31-38} \cmidrule{38-40} \cmidrule{40-42} \cmidrule{42-45} \cmidrule{45-48} \cmidrule{48-50} \cmidrule{50-53} \cmidrule{53-53}  \cmidrule{54-54}  \cmidrule{55-55} 

 & Approach & \rotatebox{90}{Reference} & \rotatebox{90}{Text} & \rotatebox{90}{Audio} & \rotatebox{90}{Image} & \rotatebox{90}{Meta (Netw.)} & \rotatebox{90}{Meta (Time-S.)} & \rotatebox{90}{Digital} & \rotatebox{90}{Analogical} & \rotatebox{90}{Explicit} & \rotatebox{90}{Implicit} & \rotatebox{90}{Matrix} & \rotatebox{90}{Support?} & \rotatebox{90}{Support?} & \rotatebox{90}{Representation} & \rotatebox{90}{Confirmatory} & \rotatebox{90}{Exploratory} & \rotatebox{90}{Predictive} & \rotatebox{90}{Content} & \rotatebox{90}{Network} & \rotatebox{90}{Meta-Data} & \rotatebox{90}{Online} & \rotatebox{90}{L./D.} & \rotatebox{90}{Ingress} & \rotatebox{90}{Analysis} & \rotatebox{90}{Abstracted} & \rotatebox{90}{2D} & \rotatebox{90}{3D} & \rotatebox{90}{S3D} & \rotatebox{90}{Select} & \rotatebox{90}{Explore} & \rotatebox{90}{Reconfigure} & \rotatebox{90}{Encode} & \rotatebox{90}{Abstract} & \rotatebox{90}{Filter} & \rotatebox{90}{Connect} & \rotatebox{90}{Direct} & \rotatebox{90}{Indirect} & \rotatebox{90}{Model Tuning} & \rotatebox{90}{Data Tuning} & \rotatebox{90}{Active Learning} & \rotatebox{90}{Iterative} & \rotatebox{90}{Progressive} & \rotatebox{90}{Numerical} & \rotatebox{90}{Textual} & \rotatebox{90}{Graphical} & \rotatebox{90}{Machine Model} & \rotatebox{90}{Mental Model} & \rotatebox{90}{Confidence} & \rotatebox{90}{Trust} & \rotatebox{90}{Privacy} & \rotatebox{90}{Types} & \rotatebox{90}{Matrix} & \rotatebox{90}{Matrix} \\

\midrule[1pt]
\multirow{10}{*}{\rotatebox[origin=c]{90}{\tiny Node-link}}

& Pajek\textsuperscript{$\dagger$}
& \cite{Batagelj.Pajek.1998}
& \symNo
& \symNo
& \symNo
& \symYes
& \symNo
& \symYes
& \symNo
& \symYes
& \symNo
& \matrixTT{black}{white}{white}{white}{white}{white}{white}{white}{white}
& \symNo
& \symNo
& \symYes
& \symYes
& \symYes
& \symNo
& \symNo
& \symYes
& \symPartial
& \symNo
& \symLetter{D}
& IIII
& II
& \symNo
& \symYes
& \symYes
& \symNo
& \symYes
& \symYes
& \symYes
& \symPartial
& \symYes
& \symYes
& \symYes
& \symYes
& \symYes
& \symNo
& \symYes
& \symNo
& \symYes
& \symYes
& \symYes
& \symNo
& \symYes
& \symNo
& \symYes
& \symPartial
& \symPartial
& \symNo
& \symExample
& \matrixDT{black}{white}{white}{white}{white}{white}
& \matrixDT{black}{white}{white}{white}{white}{white}
\\

& Gephi\textsuperscript{$\dagger$}
& \cite{Bastian.Gephi.2009}
& \symNo
& \symNo
& \symNo
& \symYes
& \symNo
& \symYes
& \symNo
& \symYes
& \symNo
& \matrixTT{black}{white}{white}{white}{white}{white}{white}{white}{white}
& \symNo
& \symNo
& \symYes
& \symYes
& \symYes
& \symNo
& \symNo
& \symYes
& \symPartial
& \symYes
& \symLetter{D}
& IIII
& II
& \symPartial
& \symYes
& \symPartial
& \symNo
& \symYes
& \symYes
& \symYes
& \symPartial
& \symYes
& \symYes
& \symYes
& \symYes
& \symYes
& \symNo
& \symYes
& \symNo
& \symYes
& \symYes
& \symYes
& \symNo
& \symYes
& \symNo
& \symYes
& \symPartial
& \symPartial
& \symNo
& \symExample
& \matrixDT{black}{white}{white}{white}{white}{white}
& \matrixDT{black}{white}{white}{white}{white}{white}
\\

& SaNDVis
& \cite{Perer.SaNDVis.2013}
& \symYes
& \symNo
& \symNo
& \symYes
& \symNo
& \symYes
& \symNo
& \symYes
& \symNo
& \matrixTT{black}{white}{white}{white}{white}{white}{white}{white}{white}
& \symNo
& \symNo
& \symYes
& \symYes
& \symYes
& \symNo
& \symYes
& \symYes
& \symNo
& \symNo
& \symLetter{D}
& IIII
& II
& \symNo
& \symYes
& \symNo
& \symNo
& \symYes
& \symYes
& \symYes
& \symNo
& \symYes
& \symYes
& \symYes
& \symYes
& \symYes
& \symNo
& \symYes
& \symNo
& \symYes
& \symYes
& \symNo
& \symNo
& \symYes
& \symNo
& \symYes
& \symNo
& \symNo
& \symNo
& \symExample\symInterview
& \matrixDT{black}{white}{white}{white}{white}{white}
& \matrixDT{black}{white}{white}{white}{white}{white}
\\

& Ghani et al.
& \cite{Ghani.VAMultimodalSNA.2013}
& \symYes
& \symNo
& \symNo
& \symYes
& \symNo
& \symYes
& \symNo
& \symYes
& \symNo
& \matrixTT{black}{white}{white}{white}{white}{white}{white}{white}{white}
& \symNo
& \symNo
& \symYes
& \symYes
& \symYes
& \symNo
& \symPartial
& \symYes
& \symYes
& \symYes
& \symLetter{D}
& II
& II
& \symNo
& \symYes
& \symNo
& \symNo
& \symYes
& \symYes
& \symYes
& \symNo
& \symYes
& \symYes
& \symYes
& \symYes
& \symYes
& \symNo
& \symYes
& \symNo
& \symYes
& \symYes
& \symNo
& \symNo
& \symYes
& \symNo
& \symYes
& \symNo
& \symNo
& \symNo
& \symInterview
& \matrixDT{black}{white}{white}{white}{white}{white}
& \matrixDT{black}{white}{white}{white}{white}{white}
\\

& Elzen et al.
& \cite{Elzen.MultivarNetwExpl.2014}
& \symPartial
& \symNo
& \symNo
& \symYes
& \symNo
& \symYes
& \symNo
& \symYes
& \symNo
& \matrixTT{black}{white}{white}{white}{white}{white}{white}{white}{white}
& \symNo
& \symNo
& \symYes
& \symYes
& \symYes
& \symNo
& \symPartial
& \symYes
& \symYes
& \symYes
& \symLetter{D}
& II
& II
& \symNo
& \symYes
& \symNo
& \symNo
& \symYes
& \symYes
& \symYes
& \symNo
& \symYes
& \symYes
& \symYes
& \symYes
& \symYes
& \symNo
& \symYes
& \symNo
& \symYes
& \symYes
& \symYes
& \symNo
& \symYes
& \symNo
& \symYes
& \symNo
& \symNo
& \symNo
& \symExample
& \matrixDT{black}{white}{white}{white}{white}{white}
& \matrixDT{black}{white}{white}{white}{white}{white}
\\

& NEREx
& \cite{El-Assady.NEREx.2017}
& \symYes
& \symNo
& \symNo
& \symNo
& \symNo
& \symYes
& \symNo
& \symYes
& \symYes
& \matrixTT{black}{black}{white}{black}{white}{white}{white}{white}{white}
& \symNo
& \symNo
& \symYes
& \symYes
& \symYes
& \symNo
& \symYes
& \symNo
& \symYes
& \symNo
& \symLetter{D}
& IIII
& II
& \symNo
& \symYes
& \symNo
& \symNo
& \symYes
& \symYes
& \symYes
& \symYes
& \symYes
& \symYes
& \symYes
& \symYes
& \symYes
& \symNo
& \symYes
& \symYes
& \symYes
& \symYes
& \symNo
& \symYes
& \symYes
& \symNo
& \symYes
& \symNo
& \symNo
& \symNo
& \symExample\symInterview
& \matrixDT{black}{white}{white}{white}{white}{white}
& \matrixDT{black}{white}{white}{white}{white}{white}
\\

& i2 Analyst's NB\textsuperscript{$\dagger$}
& \cite{IBM.AnalystsNotebook}
& \symYes
& \symPartial
& \symPartial
& \symYes
& \symPartial
& \symYes
& \symNo
& \symYes
& \symNo
& \matrixTT{black}{white}{white}{white}{white}{white}{white}{white}{white}
& \symNo
& \symNo
& \symYes
& \symYes
& \symYes
& \symYes
& \symYes
& \symYes
& \symYes
& \symYes
& \symLetter{D}
& IIII
& I
& \symYes
& \symYes
& \symNo
& \symNo
& \symYes
& \symYes
& \symYes
& \symYes
& \symYes
& \symYes
& \symYes
& \symYes
& \symYes
& \symNo
& \symYes
& \symYes
& \symYes
& \symYes
& \symYes
& \symYes
& \symYes
& \symYes
& \symYes
& \symPartial
& \symPartial
& \symPartial
& -
& \matrixDT{black}{black}{white}{black}{black}{white}
& \matrixDT{black}{black}{black}{black}{white}{white}
\\

& Palantir Gotham\textsuperscript{$\dagger$}
& \cite{Palantir.Gotham.2020}
& \symYes
& \symPartial
& \symPartial
& \symYes
& \symPartial
& \symYes
& \symNo
& \symYes
& \symNo
& \matrixTT{black}{white}{white}{white}{white}{white}{white}{white}{white}
& \symNo
& \symNo
& \symYes
& \symYes
& \symYes
& \symYes
& \symYes
& \symYes
& \symYes
& \symYes +
& \symLetter{L}\symLetter{D}
& IIII
& I
& \symYes
& \symYes
& \symNo
& \symNo
& \symYes
& \symYes
& \symYes
& \symYes
& \symYes
& \symYes
& \symYes
& \symYes
& \symYes
& \symNo
& \symYes
& \symYes
& \symYes
& \symYes
& \symYes
& \symYes
& \symYes
& \symYes
& \symYes
& \symPartial
& \symPartial
& \symPartial
& -
& \matrixDT{black}{black}{white}{black}{black}{white}
& \matrixDT{black}{black}{black}{black}{white}{white}
\\

& DataWalk\textsuperscript{$\dagger$}
& \cite{DataWalk.2020}
& \symYes
& \symPartial
& \symPartial
& \symYes
& \symPartial
& \symYes
& \symNo
& \symYes
& \symNo
& \matrixTT{black}{white}{white}{white}{white}{white}{white}{white}{white}
& \symNo
& \symNo
& \symYes
& \symYes
& \symYes
& \symYes
& \symYes
& \symYes
& \symYes
& \symYes
& \symLetter{D}
& IIII
& I
& \symYes
& \symYes
& \symNo
& \symNo
& \symYes
& \symYes
& \symYes
& \symYes
& \symYes
& \symYes
& \symYes
& \symYes
& \symYes
& \symNo
& \symYes
& \symYes
& \symYes
& \symYes
& \symYes
& \symYes
& \symYes
& \symNo
& \symYes
& \symPartial
& \symPartial
& \symPartial
& -
& \matrixDT{black}{white}{white}{white}{white}{white}
& \matrixDT{black}{white}{white}{white}{white}{white}
\\

& Nuix D./I.\textsuperscript{$\dagger$}
& \cite{Nuix.DiscoverInvestigate.2020}
& \symYes
& \symPartial
& \symPartial
& \symYes
& \symPartial
& \symYes
& \symNo
& \symYes
& \symNo
& \matrixTT{black}{white}{white}{white}{white}{white}{white}{white}{white}
& \symNo
& \symNo
& \symYes
& \symYes
& \symYes
& \symYes
& \symYes
& \symYes
& \symYes
& \symYes
& \symLetter{L}\symLetter{D}
& IIII
& II
& \symYes
& \symYes
& \symNo
& \symNo
& \symYes
& \symYes
& \symYes
& \symYes
& \symYes
& \symYes
& \symYes
& \symYes
& \symYes
& \symYes
& \symYes
& \symYes
& \symYes
& \symYes
& \symYes
& \symYes
& \symYes
& \symYes
& \symYes
& \symPartial
& \symPartial
& \symPartial
& -
& \matrixDT{black}{black}{white}{black}{black}{white}
& \matrixDT{black}{black}{black}{black}{white}{white}
\\

\midrule[1pt]
\multirow{5}{*}{\rotatebox[origin=c]{90}{\tiny Matrix}}

& Reccurence Plots
& \cite{Angus.Concept.ReccurencePlots.2012}
& \symYes
& \symNo
& \symNo
& \symNo
& \symNo
& \symYes
& \symPartial
& \symYes
& \symNo
& \matrixTT{black}{white}{white}{white}{white}{white}{white}{white}{white}
& \symNo
& \symNo
& \symYes
& \symYes
& \symYes
& \symNo
& \symYes
& \symNo
& \symNo
& \symNo
& \symLetter{D}
& II
& II
& \symNo
& \symYes
& \symNo
& \symNo
& \symYes
& \symYes
& \symYes
& \symNo
& \symNo
& \symYes
& \symYes
& \symNo
& \symYes
& \symNo
& \symNo
& \symNo
& \symNo
& \symYes
& \symNo
& \symYes
& \symYes
& \symNo
& \symYes
& \symNo
& \symNo
& \symNo
& \symExample
& \matrixDT{black}{white}{white}{white}{white}{white}
& \matrixDT{black}{white}{white}{white}{white}{white}
\\

& GestaltMatrix
& \cite{Brandes.AsymmRelSocNetw.2011}
& \symNo
& \symNo
& \symNo
& \symYes
& \symNo
& \symYes
& \symNo
& \symYes
& \symNo
& \matrixTT{black}{white}{white}{white}{white}{white}{white}{white}{white}
& \symNo
& \symNo
& \symYes
& \symYes
& \symNo
& \symNo
& \symNo
& \symYes
& \symNo
& \symNo
& \symLetter{D}
& II
& I
& \symNo
& \symYes
& \symNo
& \symNo
& \symNo
& \symYes
& \symNo
& \symNo
& \symNo
& \symNo
& \symNo
& \symNo
& \symNo
& \symNo
& \symNo
& \symNo
& \symYes
& \symNo
& \symNo
& \symNo
& \symYes
& \symNo
& \symYes
& \symNo
& \symNo
& \symNo
& \symExample\symInterview
& \matrixDT{black}{white}{white}{white}{white}{white}
& \matrixDT{black}{white}{white}{white}{white}{white}
\\

& MatrixWave
& \cite{Zhao.MatrixWave.2015}
& \symNo
& \symNo
& \symNo
& \symNo
& \symYes
& \symYes
& \symNo
& \symYes
& \symNo
& \matrixTT{black}{white}{white}{white}{white}{white}{white}{white}{white}
& \symNo
& \symNo
& \symYes
& \symYes
& \symYes
& \symNo
& \symNo
& \symNo
& \symYes
& \symNo
& \symLetter{D}
& II
& I
& \symNo
& \symYes
& \symNo
& \symNo
& \symYes
& \symYes
& \symYes
& \symNo
& \symNo
& \symYes
& \symYes
& \symYes
& \symYes
& \symNo
& \symNo
& \symNo
& \symYes
& \symYes
& \symNo
& \symNo
& \symYes
& \symNo
& \symYes
& \symNo
& \symNo
& \symNo
& \symInterview
& \matrixDT{black}{white}{white}{white}{white}{white}
& \matrixDT{black}{white}{white}{white}{white}{white}
\\

& SmallMultiPiles
& \cite{Bach.SmallMultiPiles.2015}
& \symNo
& \symNo
& \symNo
& \symYes
& \symNo
& \symYes
& \symNo
& \symYes
& \symNo
& \matrixTT{black}{white}{white}{white}{white}{white}{white}{white}{white}
& \symNo
& \symNo
& \symYes
& \symYes
& \symYes
& \symNo
& \symNo
& \symYes
& \symNo
& \symNo
& \symLetter{D}
& II
& II
& \symNo
& \symYes
& \symNo
& \symNo
& \symYes
& \symYes
& \symYes
& \symNo
& \symYes
& \symYes
& \symYes
& \symYes
& \symYes
& \symNo
& \symYes
& \symNo
& \symYes
& \symYes
& \symNo
& \symNo
& \symYes
& \symNo
& \symYes
& \symNo
& \symNo
& \symNo
& \symInterview
& \matrixDT{black}{white}{white}{white}{white}{white}
& \matrixDT{black}{white}{white}{white}{white}{white}
\\

& HyperMatrix
& \cite{Fischer.HyperMatrix.2020}
& \symYes
& \symNo
& \symNo
& \symYes
& \symNo
& \symYes
& \symNo
& \symYes
& \symNo
& \matrixTT{black}{white}{white}{black}{white}{white}{black}{white}{white}
& \symPartial
& \symNo
& \symYes
& \symYes
& \symYes
& \symYes
& \symYes
& \symYes
& \symPartial
& \symNo
& \symLetter{D}
& II
& I
& \symNo
& \symYes
& \symNo
& \symNo
& \symYes
& \symYes
& \symYes
& \symPartial
& \symYes
& \symYes
& \symYes
& \symYes
& \symYes
& \symYes
& \symYes
& \symYes
& \symYes
& \symYes
& \symYes
& \symNo
& \symYes
& \symYes
& \symYes
& \symYes
& \symYes
& \symNo
& \symExample\symComparison\symInterview
& \matrixDT{black}{black}{black}{white}{white}{white}
& \matrixDT{black}{black}{black}{white}{black}{black}
\\

\midrule[1pt]
\multirow{11}{*}{\rotatebox[origin=c]{90}{\tiny Timline}}

& Themail
& \cite{Viegas.VisEMailContent.2006}
& \symYes
& \symNo
& \symNo
& \symNo
& \symYes
& \symYes
& \symNo
& \symYes
& \symNo
& \matrixTT{black}{white}{white}{white}{white}{white}{white}{white}{white}
& \symNo
& \symNo
& \symYes
& \symYes
& \symYes
& \symNo
& \symYes
& \symNo
& \symPartial
& \symNo
& \symLetter{D}
& IIII
& II
& \symNo
& \symYes
& \symNo
& \symNo
& \symYes
& \symYes
& \symYes
& \symNo
& \symYes
& \symYes
& \symYes
& \symPartial
& \symYes
& \symNo
& \symYes
& \symNo
& \symYes
& \symYes
& \symNo
& \symYes
& \symYes
& \symNo
& \symYes
& \symNo
& \symPartial
& \symNo
& \symExample\symInterview
& \matrixDT{black}{white}{white}{white}{white}{white}
& \matrixDT{black}{white}{white}{white}{white}{white}
\\

& TextFlow
& \cite{Cui.TextFlow.2011}
& \symYes
& \symNo
& \symNo
& \symNo
& \symYes
& \symYes
& \symPartial
& \symYes
& \symNo
& \matrixTT{black}{white}{white}{white}{white}{white}{white}{white}{white}
& \symNo
& \symNo
& \symYes
& \symYes
& \symYes
& \symNo
& \symYes
& \symNo
& \symYes
& \symNo
& \symLetter{D}
& II
& II
& \symNo
& \symYes
& \symNo
& \symNo
& \symYes
& \symYes
& \symYes
& \symYes
& \symYes
& \symYes
& \symYes
& \symNo
& \symYes
& \symNo
& \symYes
& \symNo
& \symYes
& \symYes
& \symNo
& \symYes
& \symYes
& \symNo
& \symYes
& \symNo
& \symNo
& \symNo
& \symExample
& \matrixDT{black}{white}{white}{white}{white}{white}
& \matrixDT{black}{white}{white}{white}{white}{white}
\\

& CloudLines
& \cite{Krstajic.CloudLines.2011}
& \symYes
& \symNo
& \symNo
& \symNo
& \symYes
& \symYes
& \symNo
& \symYes
& \symNo
& \matrixTT{black}{white}{white}{white}{white}{white}{white}{white}{white}
& \symNo
& \symNo
& \symYes
& \symYes
& \symYes
& \symNo
& \symPartial
& \symNo
& \symYes
& \symNo
& \symLetter{D}
& IIII
& I
& \symNo
& \symYes
& \symNo
& \symNo
& \symYes
& \symYes
& \symNo
& \symNo
& \symYes
& \symYes
& \symYes
& \symNo
& \symYes
& \symNo
& \symYes
& \symNo
& \symYes
& \symNo
& \symNo
& \symNo
& \symYes
& \symNo
& \symYes
& \symNo
& \symNo
& \symNo
& \symExample
& \matrixDT{black}{white}{white}{white}{white}{white}
& \matrixDT{black}{white}{white}{white}{white}{white}
\\

& DecisionFlow
& \cite{Gotz.DecisionFlow.2014}
& \symPartial
& \symNo
& \symNo
& \symNo
& \symYes
& \symYes
& \symNo
& \symYes
& \symNo
& \matrixTT{black}{white}{white}{white}{white}{white}{white}{white}{white}
& \symNo
& \symNo
& \symYes
& \symYes
& \symYes
& \symNo
& \symPartial
& \symNo
& \symYes
& \symNo
& \symLetter{D}
& II
& II
& \symNo
& \symYes
& \symNo
& \symNo
& \symYes
& \symYes
& \symYes
& \symNo
& \symYes
& \symYes
& \symYes
& \symYes
& \symYes
& \symNo
& \symYes
& \symNo
& \symYes
& \symYes
& \symYes
& \symNo
& \symYes
& \symNo
& \symYes
& \symNo
& \symNo
& \symNo
& \symInterview
& \matrixDT{black}{white}{white}{white}{white}{white}
& \matrixDT{black}{white}{white}{white}{white}{white}
\\

& OpinionFlow
& \cite{Wu.OpinionFlow.2014}
& \symYes
& \symNo
& \symNo
& \symNo
& \symPartial
& \symYes
& \symPartial
& \symYes
& \symYes
& \matrixTT{black}{white}{white}{white}{white}{white}{white}{white}{white}
& \symNo
& \symNo
& \symYes
& \symYes
& \symYes
& \symYes
& \symYes
& \symNo
& \symPartial
& \symNo
& \symLetter{D}
& IIII
& I
& \symNo
& \symYes
& \symNo
& \symNo
& \symYes
& \symYes
& \symNo
& \symNo
& \symYes
& \symYes
& \symYes
& \symYes
& \symYes
& \symNo
& \symYes
& \symNo
& \symYes
& \symYes
& \symNo
& \symYes
& \symYes
& \symYes
& \symYes
& \symNo
& \symNo
& \symNo
& \symExample\symInterview
& \matrixDT{black}{black}{black}{white}{white}{white}
& \matrixDT{black}{black}{black}{black}{black}{black}
\\

& Han et al.
& \cite{Han.VAProxTempRel.2015}
& \symNo
& \symNo
& \symNo
& \symNo
& \symYes
& \symYes
& \symPartial
& \symPartial
& \symYes
& \matrixTT{black}{white}{white}{white}{white}{white}{white}{white}{white}
& \symNo
& \symNo
& \symYes
& \symYes
& \symYes
& \symYes
& \symPartial
& \symNo
& \symPartial
& \symNo
& \symLetter{D}
& IIII
& II
& \symNo
& \symYes
& \symNo
& \symNo
& \symYes
& \symYes
& \symYes
& \symNo
& \symYes
& \symYes
& \symYes
& \symNo
& \symYes
& \symNo
& \symYes
& \symNo
& \symYes
& \symYes
& \symNo
& \symNo
& \symYes
& \symNo
& \symYes
& \symNo
& \symNo
& \symNo
& \symExample
& \matrixDT{black}{white}{white}{white}{white}{white}
& \matrixDT{black}{white}{white}{white}{white}{white}
\\

& Liu et al.
& \cite{Liu.VisAnaTextStreams.2016}
& \symYes
& \symNo
& \symNo
& \symNo
& \symYes
& \symYes
& \symNo
& \symYes
& \symNo
& \matrixTT{black}{white}{white}{white}{white}{white}{white}{white}{white}
& \symNo
& \symNo
& \symYes
& \symYes
& \symYes
& \symNo
& \symYes
& \symNo
& \symYes
& \symNo
& \symLetter{D}
& II
& I
& \symNo
& \symYes
& \symNo
& \symNo
& \symYes
& \symYes
& \symNo
& \symNo
& \symYes
& \symYes
& \symYes
& \symNo
& \symYes
& \symNo
& \symYes
& \symNo
& \symYes
& \symYes
& \symNo
& \symNo
& \symYes
& \symNo
& \symYes
& \symNo
& \symNo
& \symNo
& \symExample
& \matrixDT{black}{white}{white}{white}{white}{white}
& \matrixDT{black}{white}{white}{white}{white}{white}
\\

& ThreadReconst.
& \cite{ElAssady.ThreadReconstructor.2018}
& \symYes
& \symNo
& \symNo
& \symNo
& \symYes
& \symYes
& \symNo
& \symYes
& \symYes
& \matrixTT{black}{white}{white}{white}{white}{white}{white}{white}{white}
& \symNo
& \symNo
& \symYes
& \symYes
& \symYes
& \symNo
& \symYes
& \symNo
& \symYes
& \symNo
& \symLetter{D}
& IIII
& II
& \symNo
& \symYes
& \symNo
& \symNo
& \symYes
& \symYes
& \symYes
& \symYes
& \symYes
& \symYes
& \symYes
& \symYes
& \symYes
& \symYes
& \symYes
& \symYes
& \symYes
& \symYes
& \symNo
& \symYes
& \symYes
& \symYes
& \symYes
& \symNo
& \symYes
& \symNo
& \symExample\symInterview
& \matrixDT{black}{white}{white}{white}{white}{white}
& \matrixDT{black}{white}{black}{white}{white}{white}
\\

& T-Cal
& \cite{Fu.TCal.2018}
& \symYes
& \symNo
& \symNo
& \symNo
& \symYes
& \symYes
& \symPartial
& \symYes
& \symNo
& \matrixTT{black}{white}{white}{white}{white}{white}{white}{white}{white}
& \symNo
& \symNo
& \symYes
& \symYes
& \symYes
& \symNo
& \symYes
& \symNo
& \symYes
& \symNo
& \symLetter{D}
& IIII
& II
& \symNo
& \symYes
& \symNo
& \symNo
& \symYes
& \symYes
& \symYes
& \symNo
& \symYes
& \symYes
& \symYes
& \symNo
& \symYes
& \symNo
& \symYes
& \symNo
& \symYes
& \symYes
& \symNo
& \symYes
& \symYes
& \symNo
& \symYes
& \symNo
& \symNo
& \symNo
& \symExample\symInterview
& \matrixDT{black}{white}{white}{white}{white}{white}
& \matrixDT{black}{white}{white}{white}{white}{white}
\\

& netflower
& \cite{Stoiber.netflower.2019}
& \symYes
& \symNo
& \symNo
& \symYes
& \symYes
& \symYes
& \symNo
& \symYes
& \symNo
& \matrixTT{black}{white}{white}{white}{white}{white}{white}{white}{white}
& \symNo
& \symNo
& \symYes
& \symYes
& \symYes
& \symNo
& \symPartial
& \symNo
& \symYes
& \symNo
& \symLetter{D}
& II
& I
& \symNo
& \symYes
& \symNo
& \symNo
& \symYes
& \symYes
& \symYes
& \symNo
& \symYes
& \symYes
& \symYes
& \symYes
& \symYes
& \symNo
& \symYes
& \symNo
& \symYes
& \symYes
& \symYes
& \symYes
& \symYes
& \symNo
& \symYes
& \symNo
& \symNo
& \symNo
& \symInterview
& \matrixDT{black}{white}{white}{white}{white}{white}
& \matrixDT{black}{white}{white}{white}{white}{white}
\\

& WeSeer
& \cite{Li.WeSeer.2020}
& \symYes
& \symNo
& \symPartial
& \symPartial
& \symYes
& \symYes
& \symPartial
& \symYes
& \symNo
& \matrixTT{black}{white}{white}{white}{white}{white}{white}{white}{white}
& \symNo
& \symNo
& \symYes
& \symYes
& \symYes
& \symYes
& \symYes
& \symPartial
& \symYes
& \symNo
& \symLetter{D}
& IIII
& II
& \symNo
& \symYes
& \symNo
& \symNo
& \symYes
& \symYes
& \symYes
& \symYes
& \symYes
& \symYes
& \symYes
& \symYes
& \symYes
& \symNo
& \symYes
& \symNo
& \symYes
& \symYes
& \symYes
& \symYes
& \symYes
& \symNo
& \symYes
& \symNo
& \symNo
& \symNo
& \symExample\symInterview
& \matrixDT{black}{black}{black}{white}{white}{white}
& \matrixDT{black}{black}{black}{white}{white}{white}
\\

\midrule[1pt]
\multirow{11}{*}{\rotatebox[origin=c]{90}{\tiny Multi-Paradigm}}

& MatrixExplorer
& \cite{Henry.MatrixExplorer.2006}
& \symYes
& \symNo
& \symNo
& \symYes
& \symNo
& \symYes
& \symNo
& \symYes
& \symNo
& \matrixTT{black}{white}{white}{white}{white}{white}{white}{white}{white}
& \symNo
& \symNo
& \symYes
& \symYes
& \symYes
& \symNo
& \symNo
& \symYes
& \symYes
& \symNo
& \symLetter{D}
& II
& I
& \symNo
& \symYes
& \symNo
& \symNo
& \symYes
& \symYes
& \symYes
& \symYes
& \symYes
& \symYes
& \symYes
& \symYes
& \symYes
& \symNo
& \symYes
& \symNo
& \symYes
& \symYes
& \symYes
& \symNo
& \symYes
& \symNo
& \symYes
& \symYes
& \symNo
& \symNo
& \symInterview
& \matrixDT{black}{white}{white}{white}{white}{white}
& \matrixDT{black}{white}{white}{black}{white}{white}
\\

& NodeTrix
& \cite{Henry.NodeTrix.2007}
& \symYes
& \symNo
& \symNo
& \symYes
& \symNo
& \symYes
& \symNo
& \symYes
& \symNo
& \matrixTT{black}{white}{white}{white}{white}{white}{white}{white}{white}
& \symNo
& \symNo
& \symYes
& \symYes
& \symYes
& \symNo
& \symNo
& \symYes
& \symYes
& \symNo
& \symLetter{D}
& II
& I
& \symNo
& \symYes
& \symNo
& \symNo
& \symYes
& \symYes
& \symYes
& \symYes
& \symYes
& \symYes
& \symYes
& \symYes
& \symYes
& \symNo
& \symYes
& \symNo
& \symYes
& \symYes
& \symYes
& \symNo
& \symYes
& \symNo
& \symYes
& \symYes
& \symNo
& \symNo
& \symExample
& \matrixDT{black}{white}{white}{white}{white}{white}
& \matrixDT{black}{white}{white}{white}{white}{white}
\\

& Hadlak et al.
& \cite{Hadlak.DynNetwClustTemp.2013}
& \symNo
& \symNo
& \symNo
& \symYes
& \symYes
& \symYes
& \symNo
& \symYes
& \symNo
& \matrixTT{black}{white}{white}{white}{white}{white}{white}{white}{white}
& \symNo
& \symNo
& \symYes
& \symYes
& \symYes
& \symNo
& \symNo
& \symYes
& \symYes
& \symNo
& \symLetter{D}
& II
& I
& \symNo
& \symYes
& \symNo
& \symNo
& \symYes
& \symYes
& \symYes
& \symYes
& \symYes
& \symYes
& \symYes
& \symPartial
& \symYes
& \symNo
& \symYes
& \symNo
& \symYes
& \symYes
& \symYes
& \symNo
& \symYes
& \symNo
& \symYes
& \symNo
& \symNo
& \symNo
& \symExample
& \matrixDT{black}{white}{white}{white}{white}{white}
& \matrixDT{black}{white}{white}{white}{white}{white}
\\

& Overview
& \cite{Brehmer.Overview.2014}
& \symYes
& \symNo
& \symNo
& \symNo
& \symYes
& \symYes
& \symNo
& \symYes
& \symNo
& \matrixTT{black}{white}{white}{white}{white}{white}{white}{white}{white}
& \symNo
& \symNo
& \symYes
& \symYes
& \symYes
& \symNo
& \symYes
& \symNo
& \symYes
& \symNo
& \symLetter{D}
& IIII
& II
& \symNo
& \symYes
& \symNo
& \symNo
& \symYes
& \symYes
& \symYes
& \symYes
& \symYes
& \symYes
& \symYes
& \symYes
& \symYes
& \symNo
& \symYes
& \symNo
& \symYes
& \symYes
& \symNo
& \symYes
& \symYes
& \symNo
& \symYes
& \symNo
& \symPartial
& \symNo
& \symExample\symInterview
& \matrixDT{black}{white}{white}{white}{white}{white}
& \matrixDT{black}{white}{white}{white}{white}{white}
\\

& ConVis
& \cite{Hoque.ConVis.2014}
& \symYes
& \symNo
& \symNo
& \symYes
& \symYes
& \symYes
& \symYes
& \symYes
& \symYes
& \matrixTT{black}{white}{white}{white}{white}{white}{white}{white}{white}
& \symYes
& \symNo
& \symYes
& \symYes
& \symYes
& \symNo
& \symYes
& \symYes
& \symYes
& \symNo
& \symLetter{D}
& II
& II
& \symNo
& \symYes
& \symNo
& \symNo
& \symYes
& \symYes
& \symYes
& \symYes
& \symYes
& \symYes
& \symYes
& \symYes
& \symYes
& \symNo
& \symYes
& \symNo
& \symYes
& \symYes
& \symYes
& \symYes
& \symYes
& \symNo
& \symYes
& \symNo
& \symNo
& \symNo
& \symInterview
& \matrixDT{black}{white}{white}{white}{white}{white}
& \matrixDT{black}{white}{white}{white}{white}{white}
\\

& TargetVue
& \cite{Cao.TargetVue.2016}
& \symYes
& \symNo
& \symPartial
& \symYes
& \symPartial
& \symYes
& \symPartial
& \symYes
& \symYes
& \matrixTT{black}{white}{white}{white}{white}{white}{white}{white}{white}
& \symNo
& \symNo
& \symYes
& \symYes
& \symYes
& \symNo
& \symYes
& \symYes
& \symYes
& \symNo
& \symLetter{D}
& II
& II
& \symNo
& \symYes
& \symNo
& \symNo
& \symYes
& \symYes
& \symYes
& \symYes
& \symYes
& \symYes
& \symYes
& \symYes
& \symYes
& \symNo
& \symYes
& \symNo
& \symYes
& \symYes
& \symYes
& \symYes
& \symYes
& \symNo
& \symYes
& \symNo
& \symNo
& \symNo
& \symExample\symInterview
& \matrixDT{black}{white}{white}{white}{white}{white}
& \matrixDT{black}{white}{white}{white}{white}{white}
\\

& MediaDiscourse
& \cite{Lu.MediaDiscourse.2016}
& \symYes
& \symNo
& \symNo
& \symNo
& \symYes
& \symYes
& \symNo
& \symYes
& \symNo
& \matrixTT{black}{white}{white}{white}{white}{white}{white}{white}{white}
& \symNo
& \symNo
& \symYes
& \symYes
& \symYes
& \symNo
& \symYes
& \symNo
& \symYes
& \symNo
& \symLetter{D}
& II
& I
& \symNo
& \symYes
& \symNo
& \symNo
& \symYes
& \symYes
& \symYes
& \symYes
& \symYes
& \symYes
& \symYes
& \symYes
& \symYes
& \symNo
& \symYes
& \symNo
& \symYes
& \symYes
& \symYes
& \symYes
& \symYes
& \symNo
& \symYes
& \symNo
& \symNo
& \symNo
& \symExample\symInterview
& \matrixDT{black}{white}{white}{white}{white}{white}
& \matrixDT{black}{white}{white}{white}{white}{white}
\\

& VisOHC
& \cite{Kwon.VisOHC.2016}
& \symYes
& \symNo
& \symNo
& \symNo
& \symYes
& \symYes
& \symPartial
& \symYes
& \symYes
& \matrixTT{black}{white}{white}{white}{white}{white}{white}{white}{white}
& \symNo
& \symNo
& \symYes
& \symYes
& \symYes
& \symNo
& \symYes
& \symNo
& \symYes
& \symNo
& \symLetter{D}
& II
& I
& \symNo
& \symYes
& \symNo
& \symNo
& \symYes
& \symYes
& \symYes
& \symYes
& \symYes
& \symYes
& \symYes
& \symYes
& \symYes
& \symYes
& \symYes
& \symNo
& \symYes
& \symYes
& \symYes
& \symYes
& \symYes
& \symNo
& \symYes
& \symYes
& \symNo
& \symNo
& \symInterview
& \matrixDT{black}{white}{white}{white}{white}{white}
& \matrixDT{black}{white}{white}{white}{white}{white}
\\

& iForum
& \cite{Fu.iForum.2017}
& \symYes
& \symNo
& \symNo
& \symYes
& \symPartial
& \symYes
& \symPartial
& \symYes
& \symYes
& \matrixTT{black}{white}{white}{black}{white}{white}{white}{white}{white}
& \symNo
& \symNo
& \symYes
& \symYes
& \symYes
& \symNo
& \symYes
& \symYes
& \symYes
& \symNo
& \symLetter{D}
& II
& II
& \symNo
& \symYes
& \symNo
& \symNo
& \symYes
& \symYes
& \symYes
& \symYes
& \symYes
& \symYes
& \symYes
& \symYes
& \symYes
& \symNo
& \symYes
& \symNo
& \symYes
& \symYes
& \symYes
& \symYes
& \symYes
& \symNo
& \symYes
& \symNo
& \symNo
& \symNo
& \symExample\symInterview
& \matrixDT{black}{white}{white}{white}{white}{white}
& \matrixDT{black}{white}{white}{white}{white}{white}
\\

& VASSL
& \cite{Khayat.VASSL.2020}
& \symYes
& \symNo
& \symPartial
& \symPartial
& \symYes
& \symYes
& \symNo
& \symYes
& \symNo
& \matrixTT{black}{white}{white}{black}{white}{white}{white}{white}{white}
& \symNo
& \symNo
& \symYes
& \symYes
& \symYes
& \symNo
& \symYes
& \symNo
& \symYes
& \symYes +
& \symLetter{D}
& II
& I
& \symNo
& \symYes
& \symNo
& \symNo
& \symYes
& \symYes
& \symYes
& \symYes
& \symYes
& \symYes
& \symYes
& \symYes
& \symYes
& \symNo
& \symYes
& \symNo
& \symYes
& \symYes
& \symYes
& \symNo
& \symYes
& \symNo
& \symYes
& \symPartial
& \symNo
& \symNo
& \symExample\symInterview
& \matrixDT{black}{white}{white}{white}{white}{white}
& \matrixDT{black}{white}{white}{white}{white}{white}
\\

& CommAID
& \cite{Fischer.CommAID.2021}
& \symYes
& \symNo
& \symPartial
& \symYes
& \symYes
& \symYes
& \symNo
& \symYes
& \symYes
& \matrixTT{black}{white}{white}{white}{white}{white}{white}{white}{white}
& \symPartial
& \symPartial
& \symYes
& \symYes
& \symYes
& \symNo
& \symYes
& \symYes
& \symYes
& \symNo
& \symLetter{D}
& II
& I
& \symPartial
& \symYes
& \symNo
& \symNo
& \symYes
& \symYes
& \symYes
& \symYes
& \symYes
& \symYes
& \symYes
& \symYes
& \symYes
& \symYes
& \symYes
& \symYes
& \symYes
& \symYes
& \symYes
& \symNo
& \symYes
& \symYes
& \symYes
& \symYes
& \symYes
& \symNo
& \symExample\symInterview
& \matrixDT{black}{white}{white}{black}{white}{white}
& \matrixDT{black}{white}{black}{black}{white}{black}
\\

\midrule[1pt]
\multirow{4}{*}{\rotatebox[origin=c]{90}{\tiny Other}}

& Zhao et al.
& \cite{Zhao.DiscourseAnalysis.2012}
& \symYes
& \symNo
& \symNo
& \symNo
& \symPartial
& \symYes
& \symPartial
& \symYes
& \symNo
& \matrixTT{black}{white}{white}{white}{white}{white}{white}{white}{white}
& \symNo
& \symNo
& \symYes
& \symYes
& \symYes
& \symNo
& \symYes
& \symNo
& \symYes
& \symNo
& \symLetter{D}
& II
& I
& \symNo
& \symYes
& \symNo
& \symNo
& \symYes
& \symYes
& \symYes
& \symYes
& \symYes
& \symYes
& \symYes
& \symYes
& \symYes
& \symNo
& \symYes
& \symNo
& \symYes
& \symYes
& \symNo
& \symYes
& \symYes
& \symNo
& \symYes
& \symYes
& \symYes
& \symNo
& \symExample\symInterview
& \matrixDT{black}{white}{white}{white}{white}{white}
& \matrixDT{black}{white}{white}{white}{white}{white}
\\

& Whisper
& \cite{Cao.Whisper.2012}
& \symYes
& \symNo
& \symPartial
& \symYes
& \symYes
& \symYes
& \symPartial
& \symYes
& \symNo
& \matrixTT{black}{black}{white}{black}{black}{white}{white}{white}{white}
& \symYes
& \symNo
& \symYes
& \symYes
& \symYes
& \symNo
& \symYes
& \symYes
& \symYes
& \symNo
& \symLetter{D}
& IIII
& I
& \symNo
& \symYes
& \symNo
& \symNo
& \symYes
& \symYes
& \symYes
& \symNo
& \symYes
& \symYes
& \symYes
& \symYes
& \symYes
& \symNo
& \symYes
& \symNo
& \symYes
& \symYes
& \symYes
& \symYes
& \symYes
& \symNo
& \symYes
& \symNo
& \symNo
& \symNo
& \symExample\symInterview
& \matrixDT{black}{white}{white}{white}{white}{white}
& \matrixDT{black}{white}{white}{white}{white}{white}
\\

& ConToVi
& \cite{El-Assady.ConToViMultiPartyTopicSpaces.2016}
& \symYes
& \symNo
& \symNo
& \symPartial
& \symNo
& \symYes
& \symNo
& \symYes
& \symYes
& \matrixTT{black}{black}{white}{black}{white}{white}{white}{white}{white}
& \symYes
& \symNo
& \symYes
& \symYes
& \symYes
& \symNo
& \symYes
& \symNo
& \symNo
& \symNo
& \symLetter{D}
& IIII
& II
& \symNo
& \symYes
& \symNo
& \symNo
& \symYes
& \symYes
& \symYes
& \symNo
& \symYes
& \symYes
& \symYes
& \symYes
& \symYes
& \symNo
& \symYes
& \symNo
& \symYes
& \symYes
& \symNo
& \symYes
& \symYes
& \symNo
& \symYes
& \symNo
& \symNo
& \symNo
& \symExample\symInterview
& \matrixDT{black}{white}{white}{white}{white}{white}
& \matrixDT{black}{white}{white}{white}{white}{white}
\\

& Beagle
& \cite{Koven.Beagle.2019}
& \symYes
& \symNo
& \symNo
& \symYes
& \symPartial
& \symYes
& \symNo
& \symYes
& \symYes
& \matrixTT{black}{white}{white}{white}{white}{white}{white}{white}{white}
& \symNo
& \symNo
& \symYes
& \symYes
& \symYes
& \symNo
& \symYes
& \symYes
& \symYes
& \symNo
& \symLetter{D}
& IIII
& II
& \symNo
& \symYes
& \symNo
& \symNo
& \symYes
& \symYes
& \symNo
& \symNo
& \symYes
& \symYes
& \symYes
& \symYes
& \symYes
& \symNo
& \symYes
& \symNo
& \symYes
& \symYes
& \symYes
& \symYes
& \symYes
& \symNo
& \symYes
& \symNo
& \symNo
& \symNo
& \symExample\symInterview
& \matrixDT{black}{white}{white}{white}{white}{white}
& \matrixDT{black}{white}{white}{white}{white}{white}
\\

		\bottomrule
	\end{tabular}%
\vspace{-0.3cm}
\end{table*}

\subsection{Visual Interface~{\color{ColorFrameworkVisualization}\inlinesymbol{symbols/icon_pillar_visualization}}} 
\label{sec:design_space_pillar_3}
While there can be many design principles involved~\cite{Conlen.DesignPrinciplesVA.2018}, we describe the visual interface abstractly~\cite{Keim.VisualAnalytics.2008}, focusing on three interrelated concepts:
\emph{representation} for the visualization, the techniques employed in \emph{interaction}, and the synthesis of both through \emph{refinement}.

\subsubsection{Representation}
The central aspect of visualization systems is their representations.

\textbf{Method}  --- %
We follow the established nomenclature of visualization techniques~\cite{Keim.VisualAnalytics.2008}.
However, we only chose those common in communication analysis:
\emph{node-link-based} (e.g.,~\cite{Batagelj.Pajek.1998}), \emph{timeline-based} (e.g.,~\cite{Gotz.DecisionFlow.2014}), and \emph{matrix-based} (e.g.,~\cite{Fischer.HyperMatrix.2020}).
\emph{Other} (e.g., chord diagrams~\cite{El-Assady.ConToViMultiPartyTopicSpaces.2016}) techniques are grouped, while we additionally highlight \emph{multiple-paradigm} (e.g., timeline, graph, and text~\cite{Fu.iForum.2017}) approaches.

\textbf{Pane}  --- %
The different visualization methods can be employed in different visualization panes.
We consider the three major ones, namely \textbf{2D}, \textbf{3D}, and \textbf{S3D} (stereoscopic 3D like VR or AR).
For example, a communication network can be visualized as a node-link diagram in either way, and each choice may influence the interaction concepts.

\subsubsection{Interaction}
Interaction methods are of central importance in visual analytics.

\textbf{Operation Method} ---  %
We classify the approaches based on their interaction method according to the classification developed by Yi et al.~\cite{Yi.InteractionInfoVis.2007}, namely \emph{Select}, \emph{Explore}, \emph{Reconfigure}, \emph{Encode}, \emph{Abstract/Elaborate}, \emph{Filter}, and \emph{Connect}.
Some are extremely common, while others like \emph{encode} depend on the capabilities.

\textbf{Manipulation}  --- %
The manipulation\cite{Keim.VisualAnalytics.2008} of the elements can be either \emph{direct}, for example, when interacting with data or visual objects.
Alternatively, it can be \emph{indirect}, for example, when modifying parameters.
Most approaches support both.

\subsubsection{Refinement}
In addition to the interaction concept, other discerning factors are the particularities of the refinement, for which we differentiate~\cite{Sacha.UncertaintyAwarenessTrust.2016} between the \emph{goal} and the \emph{strategy} to achieve it.

\textbf{Goal}  --- %
Two primary goals can be differentiated~\cite{Spinner.explAIner.2020}: is the goal to tune an underlying \emph{model} (e.g., for predicting communication behavior~\cite{Fischer.HyperMatrix.2020}) or the \emph{data} (e.g., to select a fitting representation~\cite{Liu.VisAnaTextStreams.2016})?

\textbf{Strategy} ---  %
The refinement strategy~\cite{Sacha.UncertaintyAwarenessTrust.2016} might vary: %
does it follow an \emph{iterative} (e.g., improvements through continuous interactions~\cite{El-Assady.ConToViMultiPartyTopicSpaces.2016}) or \emph{progressive} (e.g., incrementally discovering events~\cite{Krstajic.CloudLines.2011}) strategy?

\subsection{Knowledge Generation~{\color{ColorFrameworkKnowledge}\inlinesymbol{symbols/icon_pillar_knowledge}}} 
\label{sec:design_space_pillar_4}
Knowledge, generated and learned, is the ultimate analysis goal.
We propose three subcategories: \emph{output} to conceptualize the direct outcome, \emph{knowledge gain} to cover the power of the outcome, and \emph{verification approach} to consider implications and evaluations.

\subsubsection{Output}
Based on the classification of Spinner et al.~\cite{Spinner.explAIner.2020}, we propose two distinct categories for the learned knowledge type:
An \emph{explanation} can consist, for example, of numerical (e.g., graph algorithms~\cite{Bastian.Gephi.2009}), textual (e.g., presented text~\cite{Stoiber.netflower.2019}), or graphical representations (e.g., visual network representations~\cite{Brandes.AsymmRelSocNetw.2011}). It represents knowledge but in a factual representation that is not easily transferable and can be regarded as a (final) result of the existing data and is intended for humans.
In contrast to this, another type of result can be a \emph{transition function}, which is closer to an actual model, one example being the analyst's mental model.
Another type is a machine model, for example, a trained, applicable classifier (e.g., diffusion model~\cite{Wu.OpinionFlow.2014} or neural communication prediction model~\cite{Fischer.HyperMatrix.2020}) that encapsulates learned knowledge.

\subsubsection{Knowledge Gain}
As a final step in the learning process, the question arises which knowledge~\cite{Green.VAHumanCognitionModel.2008} is actually gained and how powerful the process is.

\textbf{Time Dimensionality} --- 
The time dimensionality describes the relationship between data and knowledge generation.
A 2x3 matrix shows the possible combinations of data basis and prediction type, each with the entries past, present, and future.
For example \matrixDT{black}{white}{white}{white}{white}{white} (like~\cite{Hadlak.DynNetwClustTemp.2013}), a system can use past data and predict past data, for example for a search.
Then, \matrixDT{black}{black}{black}{white}{white}{white} (like~\cite{Li.WeSeer.2020}) would be an article analysis and prediction system which has been trained on past data to analyze a text, either an existing one or one on the fly in the present and future.
Another example \matrixDT{white}{white}{white}{black}{white}{white} for a future prediction is a model that forecasts communication activity based on past events.
Note that by causality, the future is excluded.

\textbf{Predictive Power} --- 
A second consideration describes the predictive power of the knowledge generated, which is represented as a 2x3 matrix, where the result (explanation or transition function) and the time are combined.
For example \matrixDT{black}{white}{white}{white}{white}{white}, can a system explain (i.e., show) past events (virtually all systems)?
Or is it able \matrixDT{black}{black}{black}{white}{white}{white} to provide factual information for future events (e.g., information cascade prediction~\cite{Li.WeSeer.2020}, inaccessible internal model).
More powerful \matrixDT{black}{black}{black}{black}{black}{black} is a controllable model which can explain and predict (e.g., opinion diffusion~\cite{Wu.OpinionFlow.2014}).

\textbf{Domain-specific Aspects}  --- %
Here, specific options depend on the analysis tasks.
As discussed above, these are out of scope here; however, we could imagine this as future work (see Section~\ref{sec:discussion}).

\subsubsection{Verification Approach}
The presentation, as well as automated analysis of knowledge, raises a plethora of ethical as well as technical questions.

\textbf{Factors}  --- %
Visual analytics is well suited to address factors like confidence, trust \& privacy and consider aspects like fairness and accountability~\cite{Fischer.EthicalAwarenessCommAna.2022, Correll.EthicalDimensionsVis.2019}.
For example, probability scores could be used to estimate the results' confidence stemming from automatic processes (e.g., visual confidence scores~\cite{Fischer.HyperMatrix.2020}) and visualize it to the expert. 
Other examples include analysis log files, integrity protection, traceability, and verify-ability, as well as a provenance history.
The lack of such certainty measures might exclude systems from sensitive areas.
While essential, as shown by Correll~\cite{Correll.EthicalDimensionsVis.2019}, many approaches are oblivious.
For a more detailed discussion on the human factors in communication analysis, the ethical dilemmas, and design considerations for communication analysis systems, we refer to our companion work~\cite{Fischer.EthicalAwarenessCommAna.2022}.

\textbf{Evaluation}  --- %
To evaluate approaches, several options are possible:
Either \emph{examples}~\symExample or a case study (e.g., describing an application~\cite{Henry.NodeTrix.2007}).
A second option is a \emph{comparison}~\symComparison with existing approaches through feature comparison (e.g., comparing the most relevant tasks~\cite{Fischer.HyperMatrix.2020}).
A third option would be a \emph{qualitative} interview~\symInterview (e.g., interviewing eight domain experts~\cite{Fischer.CommAID.2021}) or a \emph{quantitative} user study~\symInterview.

\section{Discussion and Future Work}
\label{sec:discussion}
\label{sec:survey}
We have defined four main dimensions containing over fifty different properties, providing a conceptual framework for interactive communication analysis systems employing visual analytics principles.
This section discusses the main findings and lessons learned before reflecting on the difficulties in creating a conceptual framework for communication analysis.
In particular, we evaluate the framework through two additional domain experts and discuss the potential implications and opportunities for future research while highlighting the shortcomings. %

We imagine several \textbf{applications for this framework}:
To provide a state-of-the-art overview of the current techniques, laying the foundations for a more detailed survey. %
Further, structuring the research field and providing a common language for the community while supporting comparison between approaches for practitioners and developers alike.

\subsection{Expert Validation}
\label{sec:expert_validation}
To evaluate the relevance and validity of our conceptual framework, we recruited two additional domain experts for assessment.
Both are analytical system developers and act as consultants to domain experts in intelligence, working in law enforcement.
As such, they are highly familiar with the requirements and needs of the analysts as well as the technical implementation of the systems.
According to them, the \enquote{framework covers many use cases in intelligence}, \enquote{misses no critical dimensionality}, and is well received as helpful to structure their collection of systems.
As often \enquote{several systems are used in different stages of the investigation}, the framework is especially well suited to \enquote{assess the capabilities of presented approaches} for different tasks, to perform \enquote{market observation}, and the \enquote{basis for a detailed assessment scheme}.
The framework \enquote{can support analysts in choosing} the correct system or sequence of systems.
Currently, many investigation tasks \enquote{still require manual reading of all content} and analysis of relations, which is why more powerful and holistic approaches are needed.
Systems that \enquote{hide the complexities}, work well, and are easily understandable can \enquote{increase the analysts' performance significantly}.
Due to the ethical and legal requirements in communication analysis~\cite{Fischer.EthicalAwarenessCommAna.2022}, it becomes \enquote{increasingly relevant how explainability and information provenance} is handled and where a systematic assessment can be helpful.

\subsection{Survey Findings and Research Opportunities}
\label{sec:findings}
Our framework identifies gaps and research opportunities. %
Visual analytics is especially suited to support semi-automatic communication analysis~\cite{Herring.InteractiveMultimodalComm.2015, Jensen.CommunicationModality.2000}.
The complexity, multi-modality, and ambiguity of communication is well-suited to interactively combine domain knowledge and computing.
Concepts like interactive learning (e.g., \cite{Fischer.HyperMatrix.2020}) allow refining models, while uncertainty awareness enables automatic judgments (e.g.,~\cite{SeebacherFischer.ConversationalDynamics.2019}), fostering user trust and identifying bias~\cite{Fischer.EthicalAwarenessCommAna.2022}.

\textbf{Findings} --- To apply the framework, we have taken the 41 selected approaches (see Section~\ref{sec:methodology}) and coded them according to our conceptual framework.
Based on the results in Table~\ref{tab:survey}, we can discover several interesting aspects:
For one, studying the \textbf{data type}, while the analysis of text data seems mostly universal across representation methods, this is not the case for the other data types.
Somewhat unsurprisingly, when network data is included, the visualization is often node-link-based or multi-paradigm, while for time-series, it is either timeline or multi-paradigm.
Given the scope of the survey, the lack of audio and image is not surprising.
Given that all the approaches belong to the category of visual analytics, it is also unsurprising that virtually all support representative, confirmatory and exploratory analysis as their \textbf{analysis methodology}, their \textbf{operation methods} covered most options, and their \textbf{explanation} is at least always graphical.

More interesting, however, are the \textbf{differences} and the \textbf{research opportunities} we can conclude from their discrepancies, which we highlight in the following for each category (see Figure~\ref{fig:main_pillars}), \textbf{linking back} to \textbf{visualization systems} as well as \textbf{communication research};
aspects of particular relevance are highlighted with a star: \inlinesymbolup{symbols/icon_star}.

\begin{leftbar}{ColorFrameworkData}{ColorBGFrameworkData!50}
	\noindent\textbf{\inlinesymbolup{symbols/icon_star}~I.1 Analysis of the Meaning and Analogical Code} \newline
	Only a subset of visualization approaches analyzes the implicit meaning of the communication (e.g.,~\cite{Hoque.ConVis.2014, El-Assady.NEREx.2017, Fu.iForum.2017}).
	However, almost none analyze the \textbf{analogical code} of the communication.\newline
	\textit{\textbf{Implication:}} The analogical code can contain important cues which might support the analysis of the content and provide supportive information about the relationships inside the network, which makes it especially relevant to consider~\cite{WatzlawickBeavinJackson.Communication.1974}.
	Leveraging it can lead to a richer and more-complete analysis, while it can support resolving contradictions and ambiguities~\cite{ElAssady.ThreadReconstructor.2018}.
\end{leftbar}

\begin{leftbar}{ColorFrameworkData}{ColorBGFrameworkData!50}
	\noindent\textbf{I.2 Include Power Relations} \newline
	Again, onlx few (e.g., \cite{Cao.Whisper.2012, Hoque.ConVis.2014, El-Assady.ConToViMultiPartyTopicSpaces.2016}) approach considers the power relations between the participants, which can similarly influence the communication semantics, meaning, and modalities.\newline
	\textit{\textbf{Implication:}} Power relations between participants~\cite{SchulzvonThun.MiteinaderReden.1981, WatzlawickBeavinJackson.Communication.1974} might influence content aspects like choice of words, formality, use of irony, meaning, or meta-data aspects like dynamics~\cite{SeebacherFischer.ConversationalDynamics.2019}, timestamps, or message count.
	Results can be used and considered in context with the content analysis.
\end{leftbar}

\begin{leftbar}{ColorFrameworkData}{ColorBGFrameworkData!50}
	\noindent\textbf{I.3 Dynamic Analysis} \newline
	While some might consider this a technical problem, the development of systems that support the dynamic analysis of communication data and batch/stream approaches (e.g.,~\cite{IBM.AnalystsNotebook}) sets considerable hurdles to established analysis and visualization methods, which makes it an interesting research problem.\newline
	\textit{\textbf{Implication:}} Exploring how new data and updated results can be integrated~\cite{Cao.Whisper.2012}, how fluctuating analysis can be stabilized, and how changed predictions~\cite{Liu.VisAnaTextStreams.2016} are communicated offers more effective ways for visual communication.
\end{leftbar}

\begin{leftbar}{ColorFrameworkData}{ColorBGFrameworkData!50}
	\noindent\textbf{I.4 Research the Measurement Problem} \newline
	The measurement influence~\cite{Fischer.EthicalAwarenessCommAna.2022, Fischer.CommAID.2021} is rarely explored.\newline
	\textit{\textbf{Implication:}} Being aware of the \emph{measurement problem} and explore mitigations~\cite{Sameera.Cybercrime.2019} can strengthen user trust, while avoiding missed or erroneous results (e.g., due to codewords)~\cite{WatzlawickBeavinJackson.Communication.1974}.
\end{leftbar}

\begin{leftbar}{ColorFrameworkModel}{ColorBGFrameworkModel!50}
	\noindent\textbf{II.1 Multi-Environment Inclusion} \newline
	Many approaches lack support for data mapping and multiple data sources (e.g.,~\cite{IBM.AnalystsNotebook}), requiring preprocessed data.
	\newline
	\textit{\textbf{Implication:}} Automating the merging of heterogeneous data sources~\cite{Keim.VisualAnalytics.2008} with few or no user input reduces the amount of manual preprocessing or knowledge transfer required, make leveraging multiple data sources simultaneously less complicated~\cite{Fu.iForum.2017}, while exploring optimal interface strategies.
\end{leftbar}

\begin{leftbar}{ColorFrameworkModel}{ColorBGFrameworkModel!50}
	\noindent\textbf{\inlinesymbolup{symbols/icon_star}~II.2 Analyze Group Communication} \newline
	Only a few approaches support the analysis of group communication (e.g.,~\cite{Foltz.CommAnaTeams.2008, El-Assady.ConToViMultiPartyTopicSpaces.2016}), and almost none support nested groups.\newline
	\textit{\textbf{Implication:}} New and more detailed knowledge can be drawn how groups operate~\cite{Foltz.CommAnaTeams.2008} and information diffuses~\cite{Leavitt.CommPatternsGroup.1951} within, in particular because much communication actually happens inside or between groups, which can involve specific particularities~\cite{Bavelas.CommPatterns.1950}. 
\end{leftbar}

\begin{leftbar}{ColorFrameworkVisualization}{ColorBGFrameworkVisualization!50}
	\noindent\textbf{III.1 Visually Interactive Model Analysis} \newline
	Virtually all approaches use the 2D pane for visualizations and many automations focus on filtering instead of model tuning.\newline
	\textit{\textbf{Implication:}} Leveraging visual data analysis techniques~\cite{Keim.VisualAnalytics.2008,Yi.InteractionInfoVis.2007,Herring.InteractiveMultimodalComm.2015} and explore unused approaches like VR for improving the analysis process~\cite{Conlen.DesignPrinciplesVA.2018, Ghani.VAMultimodalSNA.2013}, focusing on the model~\cite{Spinner.explAIner.2020} instead of only data selection may allow for the higher level conclusions, supporting the knowledge generation~\cite{Sacha.KGM.2014}.
\end{leftbar}

\begin{leftbar}{ColorFrameworkKnowledge}{ColorBGFrameworkKnowledge!50}
	\noindent\textbf{\inlinesymbolup{symbols/icon_star}~IV.1 Model / Transfer Function / Knowledge Gain} \newline
	Few approaches contain an actual, powerful machine models (e.g,~\cite{Wu.OpinionFlow.2014, ElAssady.ThreadReconstructor.2018, Nuix.DiscoverInvestigate.2020, Fischer.HyperMatrix.2020, Fischer.CommAID.2021}) to analyze communication. \newline
	\textit{\textbf{Implication:}} Using such models can potentially support the analysis\cite{Green.VAHumanCognitionModel.2008, Conlen.DesignPrinciplesVA.2018}, through measures as active learning~\cite{Cao.TargetVue.2016, Ng.DLEmotionRecognition.2015}, intelligent filtering~\cite{Fischer.CommAID.2021}, or confidence-based predictions~\cite{Sacha.UncertaintyAwarenessTrust.2016}.
	Transfer Functions allow for a more universal machine learning, applying knowledge to new problems, increasing the predictive power.
	This reduces manual work while increasing analytical capabilities. %
\end{leftbar}

\begin{leftbar}{ColorFrameworkKnowledge}{ColorBGFrameworkKnowledge!50}
	\noindent\textbf{\inlinesymbolup{symbols/icon_star}~IV.2 Confidence, Trust, and Privacy} \newline
	These factors are insufficiently considered the majority of approaches, leading to a black-box analysis.
	Instead, one could include confidence estimates (e.g.,\cite{Fischer.HyperMatrix.2020}), logs, provenance (e.g.,\cite{Fischer.CommAID.2021}), data minimization, or other concepts.\newline
	\textit{\textbf{Implication:}} Several applications have strong requirements for confidence and trust~\cite{Sacha.UncertaintyAwarenessTrust.2016}, provenance~\cite{Spinner.explAIner.2020, Fischer.CommAID.2021}, and privacy~\cite{Fischer.EthicalAwarenessCommAna.2022}.
	Exploring how these can be fulfilled~\cite{Fischer.EthicalAwarenessCommAna.2022} without limiting the analysis can replace manual analysis by automated system.
\end{leftbar}

\begin{leftbar}{ColorFrameworkKnowledge}{ColorBGFrameworkKnowledge!50}
	\noindent\textbf{IV.3 Guidelines and Quantitative User Studies} \newline
	While several approaches include case studies and (qualitative) expert interviews, almost none make actual comparisons with related approaches or conduct quantitative user studies.\newline
	\textit{\textbf{Implication:}} Case studies and qualitative expert interviews are not always comparable or conducted to the same standards~\cite{Brehmer.Overview.2014}.
	While we do not doubt the systems work well as advertised, for reproducible comparisons between approaches, quantitative studies are required and evaluations along design guidelines~\cite{Conlen.DesignPrinciplesVA.2018}, providing a more objective overview.
\end{leftbar}

\begin{leftbar}{ColorInfoBoxGrayTitle}{ColorInfoBoxGrayBody}
	\noindent\textbf{\inlinesymbolup{symbols/icon_star}~O.1 Holistic Approaches} \newline
	Few approaches perform a holistic analysis (e.g., \cite{Fischer.CommAID.2021}) by considering multiple analysis aspects in context, covering all modalities.\newline
	\textit{\textbf{Implication:}} A holistic perspective~\cite{Elzen.MultivarNetwExpl.2014} can increase the analytical capabilities considerably~\cite{Brehmer.Overview.2014,Fischer.CommAID.2021}, supporting cross-matches beyond analysis boundaries~\cite{Fischer.EthicalAwarenessCommAna.2022}, while reducing mental load and manual work required.
\end{leftbar}

\begin{leftbar}{ColorInfoBoxGrayTitle}{ColorInfoBoxGrayBody}
	\noindent\textbf{\inlinesymbolup{symbols/icon_star}~O.2 Context / Analysis Reference Window}\newline
	Similar to the holistic analysis, a specific focus on the communication context in reference to each other should be explored further for both inter- and intra-modality analysis.\newline
	\textit{\textbf{Implication:}} Few approaches consider other modalities or external factors to explain particularities.
	For example, a break in a communication sequence might appear as a gap, but when combined with location information (e.g., same building) might indicate that the participants might have met for lunch and continued their conversation offline.
	In summary, the correct interpretation of communication is extremely context-dependent~\cite{Cao.TargetVue.2016, Conlen.DesignPrinciplesVA.2018}, with different applicability of analysis methods.
	Analyzing references and clues can improve the determination of the highly variable context~\cite{Mesch.SocialContextCommunicationChannels.2009} for choosing appropriate analysis methodologies.
\end{leftbar}
\vspace*{4mm}

\textbf{Implications} ---
All the previously described opportunities offer potential improvements for a more complete analysis of communication.
Fusing together multiple methods can lead to a richer and more complete analysis, potentially resolving contradictions and ambiguities.
Some opportunities might primarily support existing analysis steps (e.g., I.2) as a precursor, while others provide new areas in itself (e.g., II.2, O.1).
While the relevance of aspects might differ in any given analysis, our framework identifies and describes areas that users - and developers of interactive analysis systems - can consider and potentially leverage, depending on their analytical needs.

\subsection{Limitations and Future Work}
\label{sec:discussion_limitations_future_work}

One problem with the described taxonomy is the basis it is designed upon (\textbf{completeness}).
A significant problem in this research area is that relevant approaches are rarely labeled as belonging to communication analysis. %
We initially thought about compiling a list of domain-specific keywords for selection (e.g. social network analysis, sentiment analysis, e-mail analysis, etc.).
However, we found it highly likely that such a selection would be highly biased by our knowledge, %
which is why we decided to take a wider approach.
However, even with care it is inadvertently likely we missed individual approaches, not least by restricting the target journals and interviewing domain experts from (criminal) intelligence, although related domains like investigative journalism have similar requirements~\cite{Fischer.EthicalAwarenessCommAna.2022}. %
Also, it could happen that a few approaches fell through our automatic or manual search pattern (see Section~\ref{sec:methodology}). %
However, due to the restrictions discussed in Section~\ref{sec:methodology}, we do not claim overall completeness.
The survey forms one of three pillars for our goal of constructing the framework, and together with the other two, we are confident the majority of cases can be described within our framework.
Nevertheless, to address the issue of missing approaches, we created an accompanying \textbf{survey website} available at \href{https://communication-analysis.dbvis.de}{https://communication-analysis.dbvis.de}, which lists the approaches we considered and also allows readers to submit methods missing methods. This website is also permanently archived with an OSF repo~\cite{Fischer.OSFCommunicationSurveyWebsite.2022}.

Another possible limitation concerns the \textbf{orthogonality} of the framework itself.
Due to the complexity and heterogeneity of the area, it contains some overlaps.
As there is a need to balance the trade-off's accuracy, usability, and relevance, we think it is challenging to create a wholly consistent yet easily usable taxonomy.
The choices we made for selecting the categories are often based on the literature and justified when required.
However, given sparse taxonomy and non-standardized vocabulary, some groupings and namings could arguably have been chosen differently with the same validity.
To advance research in this area, however, we decided to propose our framework as a first possible draft and one step towards a universally accepted framework.
We, therefore, invite the research community to give feedback to stimulate the scientific discussion and extend upon the framework in further iterations, which can be enhanced further by more input from diverse research communities.
As part of this process, the individual, multi-faceted aspects can be formalized in more detail.

Another aspect is the extension of the framework to \textbf{non-human communication}.
Several aspects could be applied to communication in general, for example, machine to machine.
Indeed, nothing in the framework is specifically tailored to a human communicator. %
However, human communication is often more nuanced than machine communication, making parts of the framework less relevant, while other features (e.g., structuring, exchange content scope) might be missing so far.

\section{Conclusion}
\label{sec:conslusion}
In the last decades, communication analysis has experienced a shift away from manual analysis to computer-aided or even highly automated approaches.
However, to the same extent as automation levels increased, the analysis itself has often become more specialized, moving away from an overarching exploration.
This trend is in contradiction to traditional communication research, which stresses the importance of a holistic approach to capture the full meaning and context of communication.
As a result, many modern digital communication analysis systems are highly adapted to a narrow range of tasks, often either in the area of content \textit{or} in network analysis.
While this might be perfectly sufficient and suitable for their intended use, such an isolated analysis can sometimes lead to a less effective exploration and lead to incomplete or biased results.
Using separate approaches requires more manual work, often complicates analysis tasks, can introduce domain discontinuities, and increase the struggle domain experts face when trying to integrate their domain knowledge.
Further, an isolated analysis may not be sufficient to capture the full available information and can make the automatic as well as the manual detection of cross-matches more difficult.
The development of more holistic and advanced approaches for automated communication analysis systems is hindered by the lack of a clear framework and the absence of a common language that combines both technical aspects as well as results from traditional communication research.

We address this challenge by developing and formalizing a design space for digital communication analysis systems based on the existing tool landscape and communication research while making a case for how visual analytics principles can be employed for a more holistic approach.
By systematically discussing and structuring the different analysis areas and aspects of the design space, we arrive at a conceptual framework to provide an overview and assess the maturity of communication analysis systems.
As part of an initial survey, we have also categorized a large set of existing approaches using our framework.

By bridging the gap in the formalization of digital communication analysis systems by describing a design space for communication analysis, we aim to provide researchers with a common language, provide guidelines for building and to assess the maturity of such approaches, as well as point out gaps in the literature which offer exciting research opportunities.
The results of this work are widely applicable in a variety of domains that are concerned with communication analysis like civil security, the digital humanities, or business intelligence, both from a theoretical point of view as well as for the development of more powerful communication analysis systems.

\bibliographystyle{abbrv-doi}

\bibliography{bibliography}

\end{document}